\documentclass[journal]{IEEEtran}
\usepackage{comment}
\usepackage{multirow}
\usepackage{amsmath, amsfonts, amssymb, pifont}
\usepackage{booktabs}
\usepackage{rotating}
\usepackage{array}
\usepackage[caption=false,font=normalsize,labelfont=sf,textfont=sf]{subfig}
\usepackage{textcomp}
\usepackage{stfloats}
\usepackage{url}
\usepackage{verbatim}
\usepackage{graphicx}
\usepackage{cite}
\usepackage{romannum}
\newcolumntype{P}[2]{%
  >{\begin{turn}{#1}\begin{minipage}{#2}\footnotesize\raggedright\hspace{1pt}}c%
  <{\end{minipage}\end{turn}}%
}
\usepackage{xcolor}
\usepackage{colortbl}
\usepackage{circledsteps}
\usepackage{diagbox}
\usepackage[ruled,linesnumbered]{algorithm2e}
\usepackage{tcolorbox}
\usepackage[T1]{fontenc}
\usepackage[hidelinks]{hyperref}
\usepackage{nameref}
\usepackage{tikz}
\usetikzlibrary{shapes, arrows, positioning, backgrounds}
\definecolor{Employer}{HTML}{18AFCC}
\definecolor{Platform}{HTML}{5F9099}
\definecolor{Worker}{HTML}{CC1865}

\newcommand{\cmark}{\ding{51}}
\newcommand{\xmark}{\ding{55}}
\newcommand*{\fullref}[1]{\hyperref[{#1}]{\S\textbf{\ref*{#1}}. \nameref*{#1}}}

\begin{document}

\title{Towards Open Federated Learning Platforms: Survey and Vision from Technical and Legal Perspectives}

\author{Moming~Duan,
        Qinbin~Li,
        Linshan~Jiang,
        Bingsheng~He% <-this % stops a space
\thanks{Moming Duan, Linshan Jiang, and Bingsheng He are with the National University of Singapore, Singapore, 119077. Email:  \{\href{mailto:moming@nus.edu.sg}{moming@nus.edu.sg}, \href{mailto:linshan@nus.edu.sg}{linshan@nus.edu.sg}, \href{mailto:hebs@comp.nus.edu.sg}{hebs@comp.nus.edu.sg}\}.}%
\thanks{Qinbin Li is with the University of California, Berkeley, USA, 94720. Email: \{\href{mailto:qinbin@berkeley.edu}{qinbin@berkeley.edu}\}}
%\thanks{This paper was produced by the IEEE Publication Technology Group. They are in Piscataway, NJ.}% <-this % stops a space
%\thanks{Manuscript received April 19, 2021; revised August 16, 2021.}
}

% The paper headers
\markboth{Journal of \LaTeX\ Class Files,~Vol.~14, No.~8, August~2021}%
{Shell \MakeLowercase{\textit{et al.}}: A Sample Article Using IEEEtran.cls for IEEE Journals}

%\IEEEpubid{0000--0000/00\$00.00~\copyright~2021 IEEE}
% Remember, if you use this you must call \IEEEpubidadjcol in the second
% column for its text to clear the IEEEpubid mark.

\maketitle

\begin{abstract}
Traditional Federated Learning (FL) follows a server-dominated cooperation paradigm which narrows the application scenarios of FL and decreases the enthusiasm of data holders to participate.
To fully unleash the potential of FL, we advocate rethinking the design of current FL frameworks and extending it to a more generalized concept: Open Federated Learning Platforms, positioned as a crowdsourcing collaborative machine learning infrastructure for all Internet users. 
We propose two reciprocal cooperation frameworks to achieve this: query-based FL and contract-based FL. 
In this survey, we conduct a comprehensive review of the feasibility of constructing open FL platforms from both technical and legal perspectives.
We begin by reviewing the definition of FL and summarizing its inherent limitations, including server-client coupling, low model reusability, and non-public.
%In the query-based FL platform, which is an open model sharing and reusing platform empowered by the community for model mining, we explore a wide range of valuable topics, including the availability of up-to-date model repositories for model querying, legal compliance analysis between different model licenses, and copyright issues and intellectual property (IP) protection in model reusing.
%In the contract-based FL platform, which is an opt-in model training network that allows multi-round communication and monetization, we explain how to select training frameworks according to microtasks and discuss the pros and cons of Web3-based methods.
In particular, we introduce a novel taxonomy to streamline the analysis of model license compatibility in FL studies that involve batch model reusing methods, including combination, amalgamation, distillation, and generation. 
This taxonomy provides a feasible solution for identifying the corresponding license clauses and facilitates the analysis of potential legal implications and restrictions when reusing models.
Through this survey, we uncover the current dilemmas faced by FL and advocate for the development of sustainable open FL platforms. 
We aim to provide guidance for establishing such platforms in the future while identifying potential limitations that need to be addressed.

%Then, for query-based FL, we survey the availability of online model repositories for neural networks in the context of model query.
%Furthermore, based on mainstream model licenses, we study the legal compliance between different licenses and provide guidelines for avoiding license conflicts.

\end{abstract}


\section{Introduction}
\label{sec:intro}

%In recent years, the barriers to the development of Artificial Intelligence (AI) have been broken down with the rapid progress of ABC technologies in computing: AI, Big Data, and Cloud Computing, as well as the emergence of cost-effective specialized hardware~\cite{sze2017efficient} and software~\cite{jia2014caffe}. This has led to the world entering the third wave of AI development: Deep Learning~\cite{lecun2015deep}.
The success of current data-driven AI relies on massive amounts of training data and follows a gather-and-analyze paradigm~\cite{whang2023data}, which confronts with challenges of complying with rigorous data protection regulations such as OECD Privacy Guidelines~\cite{tene2011privacy} and GDPR~\cite{voigt2017eu}.
%Although data-centric AI is now the mainstream, a novel model-centric distributed collaborative training framework called Federated Learning is gaining popularity in both academia and industry due to its advantages in complying with privacy regulations.
Although data-centric AI is currently mainstream paradigm, Federated Learning~\cite{li2020federated}, a novel distributed collaborative training framework, is gaining popularity in both academia and industry for its advantages in complying with privacy regulations~\cite{truong2021privacy}.

\begin{figure*}[tbh]
    \centering
    \includegraphics[scale=0.9]{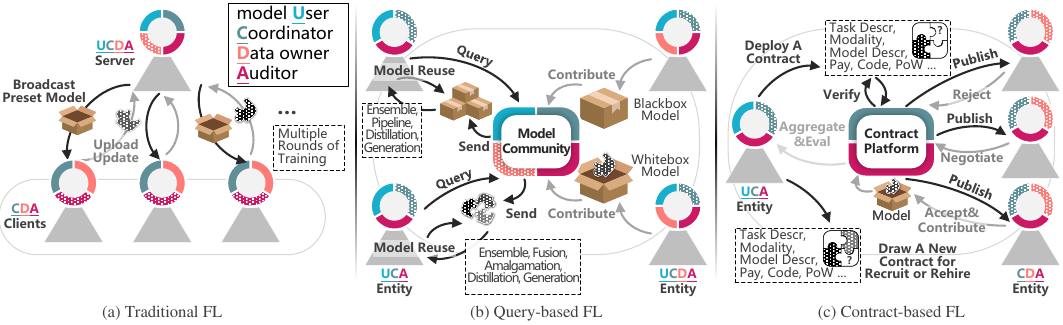} % width=\linewidth
    \caption{A schematic diagram of three cooperation frameworks of FL. (a) is the traditional FL platform, (b)~(c) are the proposed open FL platforms. Four colors correspond to four roles in~\cite{IEEEstd3652}, and colors with grid lines indicate non-essential roles.}
    \label{fig:coop}
    \vspace{-5mm}
\end{figure*}

%According to the definitions of IEEE Standard for Federated Machine Learning (FML, aka FL)~\cite{IEEEstd3652}, \textit{FL is a framework or system that enables multiple participants to collaboratively build and use machine learning models without disclosing the raw and private data owned by the participants while achieving good performance.}
%For example, a typical workflow of FL systems is that the entity with modeling demand (FL server) first deploys the FL services and initializes the training task, and then distributes this task to participants with training data (FL clients) for modeling~\cite{bonawitz2019towards}.
A typical workflow of FL systems is presented in Fig.\ref{fig:coop}(a), where the entity with a modeling demand (FL server) first deploys the FL services, initializes the training task, and then distributes this task to participants with training data (FL clients) for modeling\cite{bonawitz2019towards}.
Based on this workflow pattern, FL frameworks have been derived with specialized improvements in communication~\cite{konevcny2016federated, mcmahan2017communication, xu2023asynchronous}, optimizaiton~\cite{li2018federated, karimireddy2020scaffold, li2021model}, robustness~\cite{duan2020self, sattler2019robust, li2022federated} and privacy~\cite{bonawitz2017practical, geyer2017differentially, cheng2021secureboost}.
While these fascinating improvements greatly enhance the utility of FL, they all follow a task-based interaction paradigm, in which an FL server dominates the cooperation.
In this narrow interpretation of FL, the data owner is treated more like a worker than a collaborator and performs training primarily for the benefit of the server's goals.
Due to the above defects, FL clients have little enthusiasm to participate, and the potential for redundant training also leads to low model reusability, further diminishing the efficiency of the FL systems.
%This explains why current FL frameworks are more akin to private distributed modeling services rather than sustainable and privacy-preserving modeling platforms for everyone as expected.
This explains why current FL frameworks are more akin to private distributed modeling services rather than open and sustainable modeling platforms that every user can access and benefit from, addressing many data silo applications.

In this paper, we try to answer the question: \textbf{Can we establish a sustainable open FL platform based on a novel reciprocal cooperation framework?}
Obviously, to answer this quesion, it is insufficient to simply study the basic concepts of FL and investigate existing FL techniques.
We also need to conduct a wide survey of potential techniques that can facilitate the construction of open FL platforms.
To aid understanding, Fig.~\ref{fig:coop}(b)(c) provide a first glimpse of two novel FL cooperation frameworks that we advocated: 
\begin{itemize}
    \item \textbf{Query-based FL}. It follows a loosely-coupled cooperation framework between entities (we use "entities" instead of "clients" to emphasizes equality), where any entity can freely upload their local models or query models from an open repository named \textit{Model Community}.
    There are many valuable challenges that can be explored, such as how to query for models, how to reuse the retrieved models or how to transfer knowledge from these models, how to ensure the legal compliance between different model licenses, and how to protect the intelligent properties of released models (ref. \S\ref{sec:query}, \S\ref{sec:taxonomy} and \ddag{II}, \ddag{III}, \ddag{IV}, \ddag{V}, where \S~denotes the main section and \ddag~denotes the appendix section). %TODO: 
    \item \textbf{Contract-based FL}. It follows a mutual choice cooperation framework, where each entity can deploy model training contracts with specialized requirements such as task modality, execution environment, model architecture and license. Meanwhile, entities holding data can choose whether to accept the contract.
    We show the research topics in this setting include ML subtasks design and monetization (ref. \S\ref{sec:contract} and \ddag{VI}).
    %model pricing, model contribution evaluation and so on.% and .... (ref. Section~\ref{}) % TODO:
\end{itemize}
It is worth noting that the definitions of the four roles illustrated in Fig.~\ref{fig:coop} (i.e., model user, coordinator, data owner, auditor, ref. \S\ref{sec:basicdefinition}) are adopted for compatibility with the IEEE standard~\cite{IEEEstd3652}, and our proposals are also within the standard definitions of FL.
The diagram in Fig.~\ref{fig:coop}(a) illustrates the workflow of traditional FL, where all FL clients are required to accept the training schedule from the FL server and perform multiple rounds of local training and model averaging until the global model converges.
In contrast, the entities in query-based FL and contract-based FL are proactive in their participation.
We believe that these reciprocal cooperation frameworks have the potential to expand the prevalence of open FL and establish FL ecosystems.

\subsection{Our Contribution}
\label{sec:contribution}
  In contrast to previous surveys that primarily focused on the server-dominated cooperation framework in FL (ref. \S\ref{sec:related}),  our new survey explores the feasibility of reciprocal cooperation frameworks in FL.
  To the best of our knowledge, our work represents the first systematic survey in this area.
  The major contributions of this survey are as follows:
  \begin{itemize}
      \item We are the first to introduce the concept of open FL platforms by presenting two cooperation frameworks, namely query-based FL and contract-based FL, along with an overview of their key features and properties.
      \item We explore the query functionalities of online model repositories, such as Huggingface and OpenVINO, to investigate their feasibility for model query in query-based FL settings.
      \item We summarize the rights, restrictions, and enforcements of in-service model licenses and highlight the legal compliance and copyrightability issues in collaborative modeling. Additionally, we provide guidelines for selecting licenses to minimize conflicts and prevent license proliferation.
      \item We propose a taxonomy to streamline the legal compliance analysis in ML, which is also useful for quickly identifying suitable model reusing conceptss for open FL platforms. A comprehensive comparison of current FL studies based on this taxonomy is surveyed.
      \item We analyze the requirements for model protection in the context of query-based FL and identify applicable solutions from deep Intellectual Property (IP) protection. We also introduce the concept of designing ML microtasks for query-based FL.
\end{itemize}

\begin{comment}
The rest of this paper is organized as follows. 
We compare this survey to other related surveys and show our distinction in Section~\ref{sec:related} and Appendix~I. 
In Section~\ref{sec:basic}, we present the overview and point the limitations of traditional FL.
We comprehensively explore the feasibility and challenges of query-based FL in Section~\ref{sec:query}, which includes model query (Section~\ref{sec:how2query} and Appendix~II), model license comparison (Section~\ref{sec:licensing}) and license selection (Section~\ref{sec:choosing} and Appendix~III), copyright issues (Appendix~III.D), and analysis of license conflicts in model reusing (Appendix IV.B).
In Section~\ref{sec:taxonomy}, we present our taxonomy from a model reusing perspective, and in Appendix IV.A, we summarize FL studies based on this new taxonomy.
The discussion on model IP protection is presented in Appendix V.
We introduce how to design ML microtasks in contract-based FL in Section~\ref{sec:contract}, and we leave the discussion about decentralization and monetization to Appendix~VI.
\end{comment}

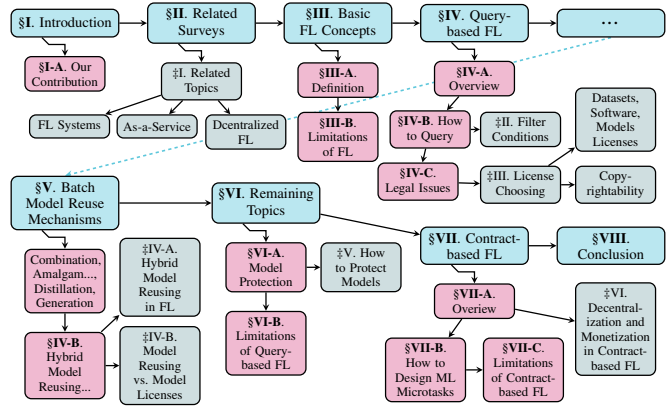
\begin{figure}[h]
    \centering
    \label{fig:article}
    \resizebox{\columnwidth}{!}{
      \begin{tikzpicture}[
        >=stealth,
        node distance=0.6cm,
        thick,
        main node/.style={rectangle, draw, fill=Employer!30, text width=2.1cm, text centered, rounded corners, minimum height=0.6cm, font=\small},
        sub node/.style={rectangle, draw, fill=Worker!30, text width=1.5cm, text centered, rounded corners, minimum height=0.5cm, font=\footnotesize},
        apdx node/.style={rectangle, draw, fill=Platform!30, text width=1.5cm, text centered, rounded corners, minimum height=0.5cm, font=\footnotesize},
        every edge/.style={draw, thick, ->}
      ]
      % Nodes
      \node[main node] (intro) {\fullref{sec:intro}};
      \node[main node, right=of intro] (related) {\fullref{sec:related}};
      \node[main node, right=of related] (concepts) {\hyperref[sec:basic]{\S\textbf{\ref{sec:basic}}. Basic FL Concepts}};
      \node[main node, right=of concepts] (qbfl) {\hyperref[sec:query]{\S\textbf{\ref{sec:query}}. Query-based FL}};
      \node[main node, right=of qbfl] (elli) {\textbf{\dots}};
      \node[main node, below=3cm of intro] (reuse) {\fullref{sec:taxonomy}};
      \node[main node, right=2cm of reuse] (remain) {\hyperref[sec:remaining_qbfl]{\S\textbf{\ref{sec:remaining_qbfl}}. Remaining Topics}};
      \node[main node, below=4cm of qbfl] (cbfl) {\hyperref[sec:contract]{\S\textbf{\ref{sec:contract}}. Contract-based FL}};
      \node[main node, right=0.6cm of cbfl] (conclusion) {\fullref{sec:conclusion}};
      
     % Arrows
     \path (intro) edge (related);
     \path (related) edge (concepts);
     \path (concepts) edge (qbfl);
     \path (qbfl) edge (elli);
     \path (reuse) edge (remain);
     \path (remain) edge (cbfl);
     \path (cbfl) edge (conclusion);
        
      % Subnodes
      % Introduction
      \node[below=0.4cm of intro, sub node] (intro_sub1) {\fullref{sec:contribution}};
      \draw[->] ([xshift=-0.5cm] intro.south) -- ++(0.15,-0.15) -|  (intro_sub1.north);

      % Related Surveys
      \node[below=0.4cm of related, apdx node] (related_sub1) {\ddag{I}. Related Topics};
      \node[below left=0.3cm and -0.8cm of related_sub1, apdx node] (related_sub2) {As-a-Service};
      \node[below right=0.3cm and -0.8cm of related_sub1, apdx node] (related_sub3) {Dcentralized FL};
      \node[below left=0.3cm and 1.1cm of related_sub1, apdx node] (related_sub4) {FL Systems};
      \draw[->] ([xshift=-0.5cm] related.south) -- ++(0.15,-0.15) -|  (related_sub1.north);
      \draw[->] (related_sub1) -- (related_sub2);
      \draw[->] (related_sub1) -- (related_sub3);
      \draw[->] (related_sub1) -- (related_sub4);
      
      % Basic Concepts of FL
      \node[below=0.4cm of concepts, sub node] (concepts_sub1) {\fullref{sec:basicdefinition}};
      \node[below=0.3cm of concepts_sub1, sub node] (concepts_sub2) {\hyperref[sec:limitations_FL]{\S\textbf{\ref{sec:limitations_FL}}. Limitations of FL}};
      \draw[->] ([xshift=-0.5cm] concepts.south) -- ++(0.15,-0.15) -|  (concepts_sub1.north);
      \draw[->] (concepts_sub1) -- (concepts_sub2);

      % Query-based FL
      \node[below=0.4cm of qbfl, sub node] (qbfl_sub1) {\fullref{sec:query_overview}};
      \node[below left=0.3cm and -0.8cm of qbfl_sub1, sub node] (qbfl_sub2) {\hyperref[sec:how2query]{\S\textbf{\ref{sec:how2query}}. How to Query}};
      \node[below left=0.3cm and -1.5cm of qbfl_sub2, sub node] (qbfl_sub3) {\hyperref[sec:how2reuse]{\S\textbf{\ref{sec:how2reuse}}. Legal Issues}};
      \node[right=0.2cm of qbfl_sub2, apdx node] (qbfl_sub4) {\ddag{II}. Filter Conditions};
      \node[below=0.4cm of qbfl_sub4, apdx node] (qbfl_sub5) {\ddag{III}. License Choosing};
      \node[right=0.3cm of qbfl_sub5, apdx node] (qbfl_sub6) {Copy-rightability};
      \node[above=0.3cm of qbfl_sub6, apdx node] (qbfl_sub7) {Datasets, Software, Models Licenses};
      \draw[->] ([xshift=-0.5cm] qbfl.south) -- ++(0.15,-0.15) -|  (qbfl_sub1.north);
      \draw[->] (qbfl_sub1) -- (qbfl_sub2);
      \draw[->] (qbfl_sub2) -- (qbfl_sub3);
      \draw[->] (qbfl_sub2) -- (qbfl_sub4);
      \draw[->] (qbfl_sub3) -- (qbfl_sub5);
      \draw[->] (qbfl_sub5) -- (qbfl_sub6);
      \draw[->] (qbfl_sub5) -- (qbfl_sub7);

      % Batch Model Reuse Mechanisms
      \node[below=0.4cm of reuse, sub node] (reuse_sub1) {\hyperref[sec:combination]{Combination}, \hyperref[sec:amalgamation]{Amalgam...}, \hyperref[sec:distillation]{Distillation}, \hyperref[sec:generation]{Generation}};
      \node[below=0.4cm of reuse_sub1, sub node] (reuse_sub2) {\hyperref[sec:how2query]{\S\textbf{\ref{sec:how2query}}. Hybrid Model Reusing...}};
      \node[right=0.3cm of reuse_sub2, apdx node] (reuse_sub3) {\ddag{IV-B}. Model Reusing vs. Model Licenses};
      \node[above=0.2cm of reuse_sub3, apdx node] (reuse_sub4) {\ddag{IV-A}. Hybrid Model Reusing in FL};
      \draw[->] ([xshift=-0.5cm] reuse.south) -- ++(0.15,-0.15) -|  (reuse_sub1.north);
      \draw[->] (reuse_sub1) -- (reuse_sub2);
      \draw[->] (reuse_sub2) -- (reuse_sub3);
      \draw[->] (reuse_sub2) -- (reuse_sub4);

      % Remaining Topics
      \node[below=0.4cm of remain, sub node] (remain_sub1) {\fullref{sec:ip_protect}};
      \node[below=0.4cm of remain_sub1, sub node] (remain_sub2) {\fullref{sec:limitations_qbfl}};
      \node[right=0.3cm of remain_sub1, apdx node] (remain_sub3) {\ddag{V}. How to Protect Models};
      \draw[->] ([xshift=-0.5cm] remain.south) -- ++(0.15,-0.15) -|  (remain_sub1.north);
      \draw[->] (remain_sub1) -- (remain_sub2);
      \draw[->] (remain_sub1) -- (remain_sub3);

      \node[below=0.4cm of cbfl, sub node] (cbfl_sub1) {\fullref{sec:contract_overview}};
      \node[below right=-0.8cm and 1.3cm of cbfl_sub1, apdx node] (cbfl_sub2) {\ddag{VI}. Decentralization and Monetization in Contract-based FL};
      \node[below left=0.4cm and -0.7cm of cbfl_sub1, sub node] (cbfl_sub3) {\fullref{sec:how2design}};
      \node[below right=0.4cm and -0.7cm of cbfl_sub1, sub node] (cbfl_sub4) {\fullref{sec:limitations_cbfl}};
      \draw[->] ([xshift=-0.5cm] cbfl.south) -- ++(0.15,-0.15) -|  (cbfl_sub1.north);
      \draw[->] (cbfl_sub1) -- (cbfl_sub2);
      \draw[->] (cbfl_sub1) -- (cbfl_sub3);
      \draw[->] (cbfl_sub3) -- (cbfl_sub4);

      \begin{scope}[on background layer]
        \draw[->, dashed, line width=1pt, dash pattern=on 2pt off 2pt, draw=Employer, opacity=0.7] (elli.south) -- (reuse.north);
      \end{scope}

      \end{tikzpicture}
    }
    \caption{Article Structure. \S: \colorbox{Employer!30}{Main Section}\colorbox{Worker!30}{Subsection}, \ddag: \colorbox{Platform!30}{Appendix}.}
    %\caption{Article Structure. \S: \textcolor{Employer!30}{Main Text Section}, \ddag: \textcolor{Platform!30}{Appendix Section}.}
    \label{fig:article_structure}
    \vspace{-5mm}
\end{figure}

The structure of this paper is shown in Fig.~\ref{fig:article_structure}. 
We compare this survey to other related surveys and show our distinction in \S\ref{sec:related}. 
In \S\ref{sec:basic}, we present the overview and point the limitations of traditional FL.
We comprehensively explore the feasibility and challenges of query-based FL in \S\ref{sec:query}, which includes model query (\S\ref{sec:how2query}), model license comparison (\S\ref{sec:licensing}) and license selection (\S\ref{sec:choosing}).
In \S\ref{sec:taxonomy}, we present our taxonomy from a model reusing perspective.
We introduce how to design ML microtasks in contract-based FL in \S\ref{sec:contract} and we conclude this paper in \S\ref{sec:conclusion}.
\emph{Due to page limits, we defer some related discussions to the appendix and we use \ddag{} to remind you that detailed analysis can be found in the appendix.}

%license selection (Section~\ref{sec:choosing} and Appendix~III), copyright issues (Appendix~III.D), and analysis of license conflicts in model reusing (Appendix IV.B).
%In Section~\ref{sec:taxonomy}, we present our taxonomy from a model reusing perspective, and in Appendix IV.A, we summarize FL studies based on this new taxonomy.
%The discussion on model IP protection is presented in Appendix V.
%We introduce how to design ML microtasks in contract-based FL in Section~\ref{sec:contract}, and we leave the discussion about decentralization and monetization to Appendix~VI.

%We conclude the paper in Section VII.

%By investigating reciprocal cooperation frameworks, we aim to broaden the understanding of FL and uncover new opportunities for collaboration and data sharing among participants. Through our comprehensive analysis, we provide valuable insights and recommendations for researchers and practitioners interested in exploring and implementing reciprocal cooperation in FL settings.

\section{Related Surveys}
\label{sec:related}
Federated learning has become a buzzword in various fields, leading to the emergence of numerous FL studies.
These works can be classified into three primary categories: FL systems design, FL applications and FL toolkits. %Extensive surveys are devoted to summarizing the advancement of federated learning, as shown in Appendix I.
The initial architectures and concepts for FL systems were summaried by Yang \textit{et al.}~\cite{yang2019federated}. 
They categorized FL into horizontal FL, vertical FL and federated transfer learning based on the distribution characteristics of data, 
which are written in IEEE Standard 3652.1-2020~\cite{yang2021white, IEEEstd3652}. 
Following this, several surveys have emerged with a focus on enhancing FL system~\cite{li2020federated,aledhari2020federated, kairouz2021advances, zhang2021survey, li2021survey}. 
From the algorithmic perspective, personlized FL~\cite{kulkarni2020survey, tan2022towards} aims to learn personlized models for each client to address the challenge of statistical heterogeneity~\cite{ma2022state}.
Meanwhile, the privacy-perserving computing platforms and model aggregation protocols for FL also been widely studied and summarized by~\cite{liu2022privacy,el2022differential,yin2021comprehensive,lyu2020threats}.
Furthermore, many advanced FL architectures had been proposed, such as asynchronous~\cite{xu2023asynchronous}, decentralized and blockchain-based FL frameworks~\cite{nguyen2021federated, qu2022blockchain, zhu2022blockchain}.
%Given that federated learning technologies enable collaboration among distributed participants in model training and decision-making, this capability holds great promise in a wide range of application scenarios.
%For instance, multiple geogrphically distributed medical insitutions can enhace medication recommendation, drug-drug interaction prediction and medical image analysis in a collaborative manner without exchanging any sensitive data~\cite{xu2021federated, pfitzner2021federated, antunes2022federated, rieke2020future}. 
%The massive real-time data generated by IoT devices in smart cities~\cite{zhang2022federated, ramu2022federated}, industries~\cite{boopalan2022fusion}, vehicles~\cite{du2020federated} has also sparked interest in exploring how FL technology can be used to deliver more advanced services such as intrusion detection, anomaly detection, fraud detection and network load prediction~\cite{agrawal2022federated, alazab2021federated, ghimire2022recent}.

Currently, most surveys extensively discuss the challenges of efficiency, heterogeneity, privacy in FL systems design, while the surveys from blockchain fields offer the most comprehensive review\textsuperscript{\ddag{I}}. 
However, except for a few blockchain-based FL studies, most of the listed surveys just present the same story from different angles and backgrounds, i.e., a server sets the model training task and delegates it to data holders to complete. 
This \textit{server-dominated} cooperation framework is a narrow implementation of FL systems.
Therefore, this survey aims to fill the gap by investigating and surveying the associated tenchnologies that support more open and inclusive cooperation frameworks in FL systems, where all entities, whether they own the data or not, can benefit from it. 

\textbf{Distinction of Our Survey}.
This survey focuses on exploring the innovative FL cooperation frameworks, which involves some FL concepts such as decentralized FL, blockchain-based FL, few-shot FL, ML related platforms and services but goes beyond them.
To the best of our knowledge, this is the first survey that focuses on the \textbf{cooperation frameworks} of FL\textsuperscript{\ddag{I}}.
%We further distinguish our survey by highlingting the similarities and differences between these related concepts in Appendix I.
%, we will differentiate this survey from other related concepts in the field of FL.

\begin{comment}
TODO:
given the high scalability of modern edge computing networks, a single MEC server cannot manage to aggregate all updates offloaded from millions of devices.
Therefore, there is an urgent need to develop a more decentralized FL approach without using a central server so as to solve security and scalability issues for enabling the next generation intelligent edge networks.
\end{comment}
% 无中心FL的性能低于传统FL

%\subsection{Blockchain-based FL}
%TODO:

%\subsection{Few-shot FL}
%TODO:

%\subsection{FAIR in FL}
%FAIR Data Principles: Findable, Accessible, Interoperable, Reusable.

\section{Basic Concepts of Federated Learning}
\label{sec:basic}
\subsection{Definition}
\label{sec:basicdefinition}
Federated Learning~\cite{IEEEstd3652, mcmahan2017communication} is a collaborative machine learning modeling paradigm that enables sharing and aggregation of knowledge from multiple sources while maintaining the confidentiality of source data.
Generally, FL systems consist of two main entities in terms of task organization: the server and the participants. 

Furthermore, FL entities can also serve multiple functional roles to support advanced features such as privacy enhancement~\cite{geyer2017differentially, bonawitz2017practical, niu2020billion}, participant scheduling~\cite{li2022federated, abdulrahman2020fedmccs}, model verification~\cite{tekgul2021waffle, shao2022fedtracker} and incentive mechanisms~\cite{yu2020fairness}.
Recall that there are four roles defined in the FL standard~\cite{IEEEstd3652}:

\begin{itemize}
    \item \textbf{Model User}. The FL model users can request for FL services and preset the targeted task, and then establish cooperation with participants who provide data. This role can leverage the benefits of collaborative training to improve the preformance of its objective models.
    
    \item \textbf{Coordinator}. The FL coordinators are responsible for providing FL services to all FL entities. This role involves setting up communication channels with entities, initializing the execution environment of participants~\cite{hanzlik2021mlcapsule}, scheduling the training and aggregation workflows to improve system efficiency.
    %, such as by alleviating the straggler effect~\cite{li2021fedsae, chai2020tifl}, optimizing data heterogeneity~\cite{duan2019astraea, abdulrahman2020fedmccs} and compressing model transfer~\cite{konevcny2016federated, sattler2019robust}.
    Additionaly, the FL coordinator provides privacy control mechanisms~\cite{bonawitz2017practical, el2022differential, hesamifard2018privacy} for model users and authorization verification for participants to maintain the security of FL systems. 
    Furthermore, the coordinator can hold a validation dataset for evaluate the models contributed by participants or detect potential disturbances from Byzantine attacks~\cite{sattler2020byzantine}.

    \item \textbf{Data Owner}. The FL data owners are knowledge contributors of FL systems, they collect and desentize raw data to maintain a local dataset for federated training. Although they have full authority over data processing and modeling, they cannot share the raw data due to privacy concerns. To address these concerns, de-identification~\cite{act1996health} and differential privacy~\cite{dwork2006differential} techniques can be applied to meet privacy budgets as required by privacy policies.
    
    \item \textbf{Auditor}. The FL auditors are responsible for formulating privacy control policies and establishing supervisory mechanisms that ensure the training process is compliant with data protection regulations (e.g., HIPAA~\cite{act1996health}, GDPR~\cite{voigt2017eu}) and preventing potential privacy breaches for both model users and data owners. Especially in FL, the latent knowledge in models can potentially reveal the sensitive information of training data~\cite{wang2019beyond, zhu2019deep, jin2021cafe}, making it crucial for auditors to scrutinize the model transmission~\cite{wei2021gradient, li2022auditing} and verify the ownership~\cite{tekgul2021waffle, shao2022fedtracker}.
\end{itemize}

\begin{figure}[t]
    \centering
    \includegraphics[width=\linewidth]{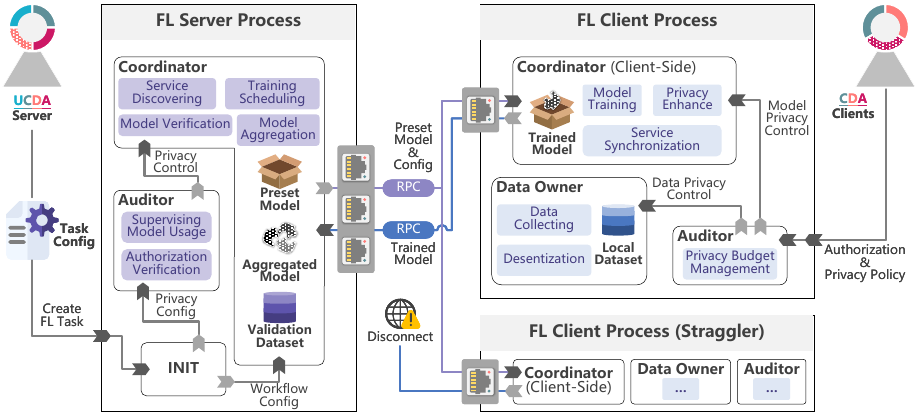} %width=\linewidth
    \caption{An overview of traditional FL systems. (U: model User, C: Coordinator, D: Data owner, A: Auditor)}
    \vspace{-5mm}
    \label{fig:fl}
  \end{figure}

Fig.~\ref{fig:fl} illustrates the typical architecture of traditional FL systems, which consists of server part and client part as a distributed modeling toolkits. 
In a general setting, the server part is the central aggregator installed in a trusted cloud environment, while the client part of software can operate in different operating environments on client devices. 
The server and clients are connected via Internet and typically with the help of Remote Procedure Call (RPC) interface for coordinating~\cite{zhang2022felicitas, abadi2016tensorflow, liu2021fate, beutel2020flower, he2020fedml, foley2022openfl}.
We use four colors to represent the four FL roles and the colors with grid lines indicate non-essential roles.
For example, in Fig.~\ref{fig:fl}, the UCDA server takes on the roles of model user, coordinator and audior. 
However, there is no necessary to hold training or validation data, so the role of data owner is non-essential.

To illustrate traditional FL workflow, we leverage the vanilla FL framework Federated Averaging (FedAvg)~\cite{mcmahan2017communication, bonawitz2019towards} as an example.
First, the FL server pre-defines the objective modeling task and initializes the server process.
Secondly,  the coordinator in server-side specifies a preset global model and the operational parameters.
Thirdly, the coordinator discovers the availability of clients' FL services, boardcasts the global model and training config to them. The training configuration contains bath size, local epoch round, optimizer parameters and so on. Then, the coordinator will wait for the trained results contributed by the coordinator in clients-side and drop those clients with network problems.
Finally, the server aggregates the trained resultes received from various clients into the global model and begins a new round based on this aggregated global model. 
The aggregation strategy adopted in FedAvg is the weighted average of model parameters based on the size of the local dataset, which means the global objective of FL can be regarded as a joint objective function of clients.
%In this way, the FL server can learn a generalized global model by jointly optimizing all local optimization objectives and incorporating the latent knowledge from the local models.
Although the auditor component was not included in prototype FedAvg, it plays an important role in the business-ready FL frameworks~\cite{liu2021fate,roth2022nvidia, ziller2021pysyft}.

However, in comparing FedAvg workflow described above with Fig.~\ref{fig:fl}, it is easy to notice that the client part has been excluded.
%This is because we are elaborating from a server-side perspective, which is usual way how FL is presented~\cite{mcmahan2017communication, li2021ditto, caldas2018leaf}.
%The underlying reason is that in traditional FL, the client-side process is tightly coupled with server-side process, and there is no alternative for clients other than to either accept or reject the training scheduling from the server wholesale.
In traditional FL, the client-side process is tightly coupled with server-side process, leaving clients to either fully accept or reject the training schedule imposed by the server with compromised audit control.
%and there is no alternative for clients other than to either accept or reject the training scheduling from the server wholesale.
So the clients are not considered as an autonomous entities but rather work as subordinates to server.
%In this server-domianted cooperation framework, the benefits and autonomy of clients are compromised, which hinders their enthusiasm to participate in FL network and subsequently limits the applicability of FL.
From this perspective, we summarize the limitations of traditional FL in the next section, which motivates us to explore more innovative sustainable FL cooperation frameworks.

\subsection{Limitations of Traditional FL}\label{sec:limitations_FL}
Previous surveys~\cite{kairouz2021advances, zhang2022federated, alazab2021federated, nguyen2021federated, zhu2022blockchain, li2020federated, yang2019federated, tan2022towards} has extensively discussed the challenges in FL systems from various aspects.
However, the cooperation mechanism of FL systems has been overlook because almost all mainsteam FL frameworks follow a same prototype~\cite{mcmahan2017communication}, which shapes the current FL form: a modeling software.
We summarize three inherent limitations of traditional FL cooperation mechanisms: (1) \textbf{Server-client Coupling}, (2) \textbf{Low Model Reusability}, (3) \textbf{Non-public}.

\subsubsection{Server-client Coupling} 
% 设备异构 从客户端层面，从服务端层面
% 不合理假设，稳定的链接和已经安装好了软件（侵入式的，后台运行，损害客户端的独立性，恶意软件，资源浪费
%The tightly-coupled server-client design is a major limitation of FL systems. From the perspective of FL service providers, adapting the programs to heterogeneous client hardware and software components, such as various operating and database systems, processor and storage architectures, communication protocols, energy constrains and data licenses, is a challenging task that significantly increases the complexity of the FL systems. 
    
%On the other hand,
The invasive software deploy mode compromises the integrity of client environments and exposes them to new privacy risks.
Specifically, the coordinator components (client-side) pushed by the server may not offer demanded privacy control mechanisms~\cite{zeng2023fedlab, caldas2018leaf, mcmahan2017communication}, or cause resource depletion on client-side~\cite{bonawitz2019towards, niu2020billion, chen2020deep}, or even piggyback malicious executable codes~\cite{li2017understanding}.
So the auditor role of client is non-essential as depicted in Fig.~\ref{fig:fl}, not only because the client maybe lacks a corresponding policy for FL training, but also because its privacy is not completely under its control.
Likewise, the malicious clients can also exploit the vulnerability in the aggreagation strategy to currupt the FL tranining process~\cite{bouacida2021vulnerabilities, sattler2020byzantine, park2021sageflow, fang2020local} or insert backdoors~\cite{bagdasaryan2020backdoor, wang2020attack}.
In addition, the unstable network environment can drive clients to drop out from traning (i.e., straggler effect), thereby reducing system efficiency~\cite{reisizadeh2019robust, park2021sageflow}.
Therefore, the server-client coupling design of traditional FL systems make them susceptible to unpredictable runtime environments, leading to system vulnerability and low reliability.

\subsubsection{Low Model Reusability} % 模型的移植性 不可持续 模型的可发现性
The traditional FL scheduling follows a task-centric manner and terminates once the training reaches a preset number of rounds or meets traget metrics on global model set by FL server~\cite{bonawitz2019towards}.
As a result, only FL server can guarantee having the latest global model after the task is terminated.
This ad-hoc modeling paradigm results in low model reusability and transportability.
%For example, if a client who participated in the previous training turn wants to continue, they can only start the task from scratch unless they have the up-to-date global model.
Since only FL server is able to maintain the complete modeling trajectory, it is difficult for the client to roll back the training itself to eliminate the potential privacy risk.
Meanwhile, the non-deliverable tasks scheduling mechanism also hinders inter-task model reuse.%, which leads to unnecessary energy waste and time cost on participants that have been involved in similar tasks.

\subsubsection{Non-public} % 无法自由发起任务，不提供公共接口
Except PySyft~\cite{ziller2021pysyft}, the application scenarios of mainstream FL frameworks~\cite{liu2021fate, abadi2016tensorflow, zeng2023fedlab, caldas2018leaf, ibmfl2020ibm, he2020fedml, beutel2020flower, roth2022nvidia} aim to provide private collaborative ML traning service, and there is no any accessible FL platform for the public\textsuperscript{\ddag{I-A}}.
Although there have been real-world deployment practices of FL for the public with scales of millions~\cite{bonawitz2019towards} and billions~\cite{niu2020billion}, these have been carried out only by tech giants with a massive base of active users. For an individual user, there is no practical way to organize such a large-scale FL training network.

Due to the limitations in the cooperation mechanism mentioned above, data owners are not sufficiently motivated to participate in this server-take-all FL training network even if it is public accessible. 
Therefore, the cornerstone of buliding a sustainable open FL platform is to create a reciprocal FL cooperation framework, followed by corresponding multi-source knowledge aggregation strategies, which we survey through the following two themes:

In the query-based FL platform, which is an open model sharing and reusing platform empowered by the community for model mining, we explore a wide range of valuable topics, including the availability of up-to-date model repositories for model querying, legal compliance analysis between different model licenses, and copyright issues and IP protection in model reusing.
In the contract-based FL platform, which is an opt-in model training network that allows multi-round communication and monetization, we explain how to select training frameworks according to microtasks and discuss the pros and cons of Web3-based methods.
We will detail discuss these topics in the following sections.

\section{Query-based Federated Learning}
\label{sec:query}
\subsection{Overview}
\label{sec:query_overview}
\begin{figure}[t]
  \centering
  \includegraphics[width=\linewidth]{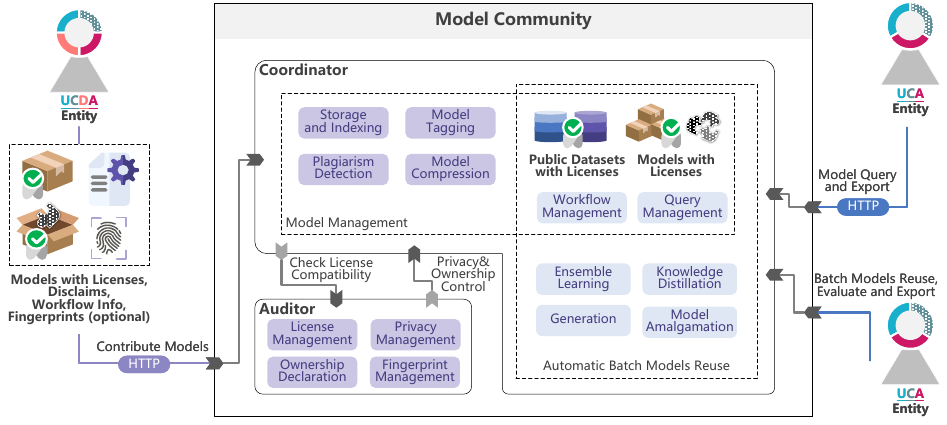}
  \caption{An overview of query-based FL systems. (U: model User, C: Coordinator, D: Data owner, A: Auditor)}
  \vspace{-5mm}
  \label{fig:query}
\end{figure}

Let us continue by establishing a sustainable open FL platform based on a query-based cooperation framework.
An overview of this platform is presented in Fig.~\ref{fig:query}, the desin philosophy behind this framework is to break the coupling between FL server and clients.
In the query-based FL systems, all traditional FL roles and components are maintained on an open model repository called Model Community. The Model Community privdes a one-stop ML models redistribution and reuse service, including model indexing, automatic batch model reuse, license management, privacy control and so on.
In addition to large-scale pretrained models like BERT~\cite{devlin2018bert}, BLOOM~\cite{scao2022bloom} with great generalization abilities, we also encourage individuals to upload their task-specific models trained on limited domain data to boost the knowledge mining within models, aka model mining~\cite{you2021workshop}.
The derivatives of model mining can learn representations from multiple domains, resulting in more promising performance that can be evaluated by platform users.
Furthermore, the contributors can release models under applicable licenses, granting them distribution control and legal protection of their intellectual property (IP).
In summary, the properties of query-based FL are:
(1) \textbf{Model Agnostic}, as there are no restrictions on the types and architectures of the models uploaded by users;
(2) \textbf{Contactless}, as communication channels need not be maintained; 
(3) \textbf{Community-powered}, whereby sharing models enrichs the entire community.

Actually, we aim to advocate a novel Software-as-a-Service (SaaS)~\cite{brereton1999future} ML platform with automatic batch model reuse integrated, which has potential to leverage the transportability of models to address previously unexplored ML problems.
Due to the high computational demands of deep learning, current ML platforms primarily concentrate on computing, for example, MaaS, MLaaS, FLaaS provide ML models deployment and development services to handle user-specified tasks\textsuperscript{\ddag{I-B}}.
On the other hand, there are several ML platforms provide open model search and download services. 
So, can we leverage leverage off-the-shelf public model repositories to build a query-based FL system?
Unfortunately, these sites are designed solely for sharing and are not suitable for more advanced functionalities such as model ensemble~\cite{jacobs1991adaptive} and knowledge distillation (KD)~\cite{hinton2015distilling}, and we will explain the reasons in the following section.

% Model Mining

\subsection{How to Query for Models}
\label{sec:how2query}
To establish a query-based FL platform, the first thing that comes to mind is how to query for models.
Unlike traditional ML model sharing repositories that mainly query for a specific model by name, it requires an efficiency approach to export a batch of target models that are ready for ensemble or distillation.
We summarized the filter conditions of existing DNNs sharing repositories\footnote{Hugging Face: \url{https://huggingface.co}; Model Zoo: \url{https://modelzoo.co}; Tensorflow Hub: \url{https://tfhub.dev}; NVIDIA NGC: \url{https://catalog.ngc.nvidia.com/models};  OpenVINO: \url{https://docs.openvino.ai/latest/model\_zoo.html}; Pytorch Hub: \url{https://pytorch.org/hub}} in TABLE~\ref{table:repository}.
The prevailing method for querying models involves searching for the desired model by its name, used datasets, and the associated tasks.
For instance, one might search for the model name GPT~\cite{radford2019language}, models trained on the MNIST dataset~\cite{lecun2010mnist}, or models capable of performing image segmentation tasks.
However, this model retrieval method requires the users have a strong priori knowledge in data science, thus raising the barrier for knowledge mining within models.
As an example, there is no effective way to acquire a batch of image classfication models that contains the knowledge of \textit{lesser panda} for further distillation.
A compromise solution is to manually search the schema of each dataset one-by-one and subsequently search for models trained on those datasets.

As shown in TABLE~\ref{table:repository}, most DNNs repositories are simply listing the description of input/output (e.g., NVIDIA NGC, OpenVINO) or even just present the source codes (e.g., Tensorflow Hub, Pytorch Hub).
This lack of unified convention for model input/output poses a challenge for query-based FL.
Additionally, most of DNNs repositories do not enable querying models by licenses, resulting in the cumbersome task of individually handling model licenses and ensuring compatibility among different licenses.
Hence, it is imperative to reconsider the design of DNNs repositories to enable quick identification of readily reusable models for knowledge aggregation. 
We further suggest following filter conditions for query-based FL: 1) Data Description; 2) Workflow and History; 3) Software Dependency; 4) Fairness and Robustness\textsuperscript{\ddag{II}}. %We leave the detailed reasons in Appendix~II.

%%% FACTsheet

\begin{table*}[t]
  \centering
  \caption{Filter conditions and characteristics of DNNs repositories. \cmark : Supported, \xmark : Unsupported, \textbf{!} : Information provided but unsearchable, listed in descending order by number of released models. (Accessed on January 17, 2024)}
  \label{table:repository}
  \footnotesize
  \begin{tabular}{|l|c|c|c|c|c|c|c|c|}
  \hline
  & \multicolumn{1}{l|}{DS Name} & \multicolumn{1}{l|}{Model Architecture} & \multicolumn{1}{l|}{Modality/Task} & \multicolumn{1}{l|}{Tag} & \multicolumn{1}{l|}{License} & \multicolumn{1}{l|}{Input-Output} & \multicolumn{1}{l|}{Batch Export} & \multicolumn{1}{l|}{\# of Models}\\ \hline
  Hugging Face & \cmark & \cmark & \cmark & \cmark & \cmark & \textbf{!} & \xmark & 470,263 \\ \hline
  Model Zoo & \cmark & \cmark & \cmark & \cmark & \xmark & \xmark & \xmark & 3,245 \\ \hline
  Tensorflow Hub & \cmark & \cmark & \cmark & \cmark & \textbf{!} & \textbf{!} & \xmark & 2,186 \\ \hline
  NVIDIA NGC & \textbf{!} & \cmark & \cmark & \cmark & \textbf{!} & \textbf{!} & \xmark & 680 \\ \hline
  OpenVINO & \textbf{!} & \cmark & \cmark & \xmark & \textbf{!} & \textbf{!} & \cmark & 277 \\ \hline
  Pytorch Hub & \textbf{!} & \cmark & \xmark & \xmark & \xmark & \textbf{!} & \xmark & 52 \\ \hline
  \end{tabular}
  \vspace{-5mm}
\end{table*}

The aforementioned filter conditions provide comprehensive coverage of the ML modeling process. 
However, there are additional requirements depending on the reuse mechanisms in the model retrieval side. 
For example, FedAvg~\cite{mcmahan2017communication} aggregates the local models weights element-wise, which requires full access to the models. 
In contrast, MoE~\cite{jacobs1991adaptive} only ensembles a batch of model outputs, so the individual models can remain blackboxes in this scenario.
So, in the context of software licenses or model licenses, the batch models reused by FedAvg should be released as source code, while those reused by MoE can be released as binary executable modules (e.g., dynamic linking).
The above distinction is crucial for ensuring that model reuse results meet the legal framework, and this has been overlooked in traditional FL.
We will expand on this topic in the following section.

\subsection{Legal Considerations in Batch Model Reuse}
\label{sec:how2reuse}
Once we have acquired a batch of models that can contribute to the new target task, the next step is to reuse the knowledge of these pre-trained models, i.e., transfer their knowledge from source domain to the target domain~\cite{pan2009survey}.
However, before deciding on how to reuse the model, it is important to ensure that the rights and permissions have been obtained. 
This may involve reviewing the terms and conditions of the licenses under which the models were released or obtaining permission from the original creators or copyright holders.
Therefore, in this section, we will not focus on the technical details of how to reuse models, which is already covered by many related surveys, such as Transfer Learning~\cite{pan2009survey}, Ensemble Leanring~\cite{zhou2012ensemble}, Domain Adaptation~\cite{wang2018deep}, Knowledge Distillation~\cite{wang2021knowledge}, Deep Generative Models~\cite{cao2022survey} and Model Fusion~\cite{ji2021emerging}.
Meanwhile, the specific model reuse techniques used are at the user's discretion, and the query-based FL platform is not bounded or restricted to any particular reuse method.
Innovatively, we study how to reuse batch of models, from the perspective of \textbf{legal compliance}.
%Therefore, the focus of this section is how to reuse batch of models, from the perspective of legal compliance.

The machine learning community benefits from the openness of ideas and code, and many high-impact ML conferences and journals encourage authors to publish their source code and dataset to research platforms like Papers With Code and Code Ocean\footnote{\url{https://paperswithcode.com}; \url{https://codeocean.com}} to increase exposure and facilitate reproducibility.
To restrict the use of ML techniques for unethical purposes (i.e., Deepfakes~\cite{mirsky2021creation}) and protect the IP of creators, models are typically published under a license agreed upon by the licensor.
Here, we summarized the granted rights, restrictions and enforcements of licenses for ML models posted on Hugging Face in TABLE~\ref{tab:licenses}.
The following sections will provide a detailed survey of these licenses.

\begin{table*}[t]
  \centering
  \scriptsize
  \caption{Licenses for ML models available on Hugging Face with a focus on their rights, restrictions and enforcements, grouped by FOSS licenses, AI model licenses, free content or database licenses in descending order of number of models (GPL, BSD, LGPL, CC licenses with unspecified versions are excluded, the similar revisions are merged). \cmark : Permited or Required, \xmark : Not Permited or Not Required, \textbf{!} : Not Explicitly Permited, * : Copyleft License, $^{\dagger}$ : Public Domain License. Only the source code of the original work under AFL-3.0 or Artistic-2.0 is required to be disclosed. You may not distribute the modified materials licensed under CC-BY-NC-ND or CC-BY-ND. (Accessed on January 17, 2024)}
  \label{tab:licenses}
  \begin{tabular}{r||ccc|ccc|cccc|c|p{3.6cm}}
    \toprule
    Licenses
    & \multicolumn{1}{P{90}{2.0cm}}{Modify / Merge} &
      \multicolumn{1}{P{90}{2.0cm}}{Redistribution} &
      \multicolumn{1}{P{90}{2.0cm}}{Sublicensing} & 
      \multicolumn{1}{P{90}{2.0cm}}{Commercial Use} & 
      \multicolumn{1}{P{90}{2.0cm}}{Patent Use} & 
      \multicolumn{1}{P{90}{2.0cm}}{Trademark Use} &
      \multicolumn{1}{P{90}{2.0cm}}{State Changes} &
      \multicolumn{1}{P{90}{2.0cm}}{Disclose Source} &
      \multicolumn{1}{P{90}{2.0cm}}{Responsible-use Restrictions} &
      \multicolumn{1}{P{90}{2.3cm}}{License/Attribution Preservation} &
      \multicolumn{1}{P{90}{2.0cm}}{\# of Models} &
      \multicolumn{1}{c}{Licensed Materials / Remarks}    \\
    \midrule

    \rowcolor{green!15}
    Apache-2.0 & \cmark & \cmark & \cmark & \cmark & \cmark & \xmark & \cmark & \xmark & \xmark & \cmark & 65,985 & BERT~\cite{devlin2018bert} \\

    MIT &  \cmark & \cmark & \cmark & \cmark & \textbf{!} & \textbf{!} & \xmark & \xmark & \xmark & \cmark & 30,344 & GPT-2~\cite{radford2019language} \\

    \rowcolor{green!15}
    AFL-3.0 & \cmark & \cmark & \cmark & \cmark & \cmark & \xmark & \cmark & (\cmark) & \xmark & \cmark & 2,208 & Italian-Legal-BERT~\cite{licari2022italian} \\

    % LLama2 & \cmark & \cmark & \cmark & \cmark & \cmark & \xmark & \cmark & (\cmark) & \xmark & \cmark & 1,342 & LLama2 \\

    *GPL-3.0 & \cmark & \cmark & \xmark & \cmark & \cmark & \xmark & \cmark & \cmark & \xmark & \cmark & 1,242 & PersonaGPT~\cite{tang2021persona} \\

    \rowcolor{green!15}
    Artistic-2.0 & \cmark & \cmark & \cmark & \cmark & \cmark & \xmark & \cmark & (\cmark) & \xmark & \cmark & 675 & Include original source \\

    %606+30
    BSD-3-Clause\&-Clear & \cmark & \cmark & \cmark & \cmark & \textbf{!} & \xmark & \xmark & \xmark & \xmark & \cmark & 636 & CodeGen~\cite{nijkamp2023codegen}/ A MIT-style license \\

    \rowcolor{green!15}
    $^{\dagger}$WTFPL-2.0 & \cmark & \cmark & \textbf{!} & \cmark & \textbf{!} & \textbf{!} & \xmark & \xmark & \xmark & \xmark & 409 & A MIT-style permissive license  \\

    *AGPL-3.0 & \cmark & \cmark & \xmark & \cmark & \cmark & \xmark & \cmark & \cmark & \xmark & \cmark & 265 & Extended GPL covers SaaS  \\

    \rowcolor{green!15}
    $^{\dagger}$Unlicense & \cmark & \cmark & \textbf{!} & \cmark & \textbf{!} & \textbf{!} & \xmark & \xmark & \xmark & \xmark & 254 & A MIT-style permissive license  \\
    %GPL & 1 & 2 & 3 & 4 & 5 & 6 & 7 & 8 & 9 & 10 & 63 &  \\

    *GPL-2.0 & \cmark & \cmark & \xmark & \cmark & \textbf{!} & \textbf{!} & \cmark & \cmark & \xmark & \cmark & 91 & Not compatible with GPL-3.0  \\

    %77+7
    \rowcolor{green!15}
    *LGPL-3.0\&2.1 & \cmark & \cmark & \xmark & \cmark & \cmark & \textbf{!} & \cmark & \cmark & \xmark & \cmark & 84 & For software libraries  \\

    BSD-2-Clause & \cmark & \cmark & \cmark & \cmark & \textbf{!} & \textbf{!} & \xmark & \xmark & \xmark & \cmark & 82 & A MIT-style permissive license  \\

    \rowcolor{green!15}
    BSL-1.0 & \cmark & \cmark & \cmark & \cmark & \textbf{!} & \textbf{!} & \xmark & \xmark & \xmark & \cmark & 77 & A MIT-style permissive license \\
    %BSD & 1 & 2 & 3 & 4 & 5 & 6 & 7 & 8 & 9 & 10 & 43 &  \\

    *OSL-3.0 & \cmark & \cmark & \cmark & \cmark & \cmark & \xmark & \cmark & \cmark & \xmark & \cmark & 55 & Linking is not derivative work \\

    \rowcolor{green!15}
    *Ms-PL & \cmark & \cmark & \cmark & \cmark & \cmark & \xmark & \xmark & \xmark & \xmark & \cmark & 43 & Weak copyleft license \\ % Weak copyleft

    %BSD-3-Clause-Clear & 1 & 2 & 3 & 4 & 5 & 6 & 7 & 8 & 9 & 10 & 14 &  \\
    %LGPL & 1 & 2 & 3 & 4 & 5 & 6 & 7 & 8 & 9 & 10 & 12 &  \\
    ECL-2.0 & \cmark & \cmark & \cmark & \cmark & \cmark & \xmark & \cmark & \xmark & \xmark & \cmark & 38 & For education communities \\

    \rowcolor{green!15}
    Zlib & \cmark & \cmark & \textbf{!} & \cmark & \textbf{!} & \textbf{!} & \xmark & \xmark & \xmark & \cmark & 30 & Rename if modified \\

    *MPL-2.0 & \cmark & \cmark & \cmark & \cmark & \cmark & \xmark & \cmark & \cmark & \xmark & \cmark & 23 & State changes under MPL only  \\

    %14+5
    \rowcolor{green!15}
    *EPL-2.0\&1.0 & \cmark & \cmark & \cmark & \cmark & \cmark & \textbf{!} & \xmark & \cmark & \xmark & \cmark & 19 & Can link proprietary license code \\

    ISC & \cmark & \cmark & \textbf{!} & \cmark & \textbf{!} & \textbf{!} & \xmark & \xmark & \xmark & \cmark & 15 & MIT-style license w/o sublicense \\ % Permission to distribute this software for any purpose

    \rowcolor{green!15}
    *EUPL-1.1 & \cmark & \cmark & \cmark & \cmark & \cmark & \xmark & \cmark & \cmark & \xmark & \cmark & 15 & License of EU covers SaaS \\

    NCSA & \cmark & \cmark & \cmark & \cmark & \textbf{!} & \xmark & \xmark & \xmark & \xmark & \cmark & 10 & Include full text of license \\

    \rowcolor{green!15}
    PostgreSQL & \cmark & \cmark & \textbf{!} & \cmark & \textbf{!} & \textbf{!} & \xmark & \xmark & \xmark & \cmark & 7 & A MIT-style license \\

    %\rowcolor{green!15}
    %OFL-1.1 & \cmark & \cmark & \xmark & \cmark & \textbf{!} & \textbf{!} & \xmark & \xmark & \xmark & \cmark & 3 & For font software \\

    %EPL-1.0 & 1 & 2 & 3 & 4 & 5 & 6 & 7 & 8 & 9 & 10 & 2 &  \\
    %LGPL-2.1 & 1 & 2 & 3 & 4 & 5 & 6 & 7 & 8 & 9 & 10 & 1 &  \\

    %\rowcolor{green!15}
    %LPPL-1.3c & \cmark & \cmark & \cmark & \cmark & \textbf{!} & \xmark & \cmark & \cmark & \xmark & \cmark & 4 & Covering  stewardship transfer \\
    
    \hline
    %\textbf{Model Licenses $\downarrow$} & \multicolumn{12}{l}{} \\
    \hline
    
    OpenRAIL &  \multicolumn{10}{l|}{>Responsible AI License template, w/o full text} & 22,947 & ControlNet~\cite{zhang2023adding}  \\

    \rowcolor{yellow!15}
    CreativeML-OpenRAIL-M & \cmark & \cmark & \cmark & \cmark & \cmark & \xmark & \cmark & \xmark & \cmark & \cmark & 15,591 & Stable Diffusions v1~\cite{rombach2022high} \\

    Llama2 & \cmark & \cmark & \xmark & (\cmark) & \cmark & \xmark & \xmark & \xmark & \cmark & \cmark & 3,538 & Llama 2~\cite{touvron2023llama} \\

    \rowcolor{yellow!15}
    OpenRAIL++ & \multicolumn{10}{l|}{>Same as CreativeML-OpenRAIL-M} & 1,433 & Stable Diffusion v2~\cite{rombach2022high} \\

    BigScience-OpenRAIL-M & \multicolumn{10}{l|}{>Same as BigScience-BLOOM-RAIL-1.0} & 659 & A general version of 1.0 \\

    \rowcolor{yellow!15}
    BigScience-BLOOM-RAIL-1.0 & \cmark & \cmark & \cmark & \cmark & \cmark & \xmark & \cmark & \xmark & \cmark & \cmark & 527 & BLOOM~\cite{scao2022bloom} \\

    BigCode-OpenRAIL-M & \multicolumn{10}{l|}{>Same as BigScience-BLOOM-RAIL-1.0} & 320 & StarCoder~\cite{li2023starcoder} \\

    \rowcolor{yellow!15}
    OPT-175B & \cmark & \xmark & \xmark & \xmark & \xmark & \xmark & \xmark & \xmark & \cmark & \cmark & $\approx94$ & OPT LLM~\cite{zhang2022opt} \\

    SEER &  \multicolumn{10}{l|}{>Same as OPT-175B, ban on reverse-engineer} & $\approx23$ & SEER Vision Model~\cite{goyal2022vision} \\
    
    \hline
    \hline

    %4,602+57+88
    \rowcolor{blue!15} 
    CC-BY-NC-4.0\&3.0\&2.0 & \cmark & \cmark & \xmark & \xmark & \xmark & \xmark & \cmark & \xmark & \xmark & \cmark & 4,747 & GALACTICA~\cite{taylor2022galactica} \\

    % 3,201+114+97+17
    CC-BY-4.0\&3.0\&2.5\&2.0 & \cmark & \cmark & \xmark & \cmark & \xmark & \xmark & \cmark & \xmark & \xmark & \cmark & 3,429 & RoBERTa-SQuAD2.0~\cite{rajpurkar2016squad} \\

    %1,739+26+18
    \rowcolor{blue!15}
    *CC-BY-NC-SA-4.0\&3.0\&2.0 & \cmark & \cmark & \xmark & \xmark & \xmark & \xmark & \cmark & \cmark & \xmark & \cmark & 1,783 & LayoutLMv3~\cite{huang2022layoutlmv3} \\

    % 1,209+301
    *CC-BY-SA-4.0\&3.0 &  \cmark & \cmark & \xmark & \cmark & \xmark & \xmark & \cmark & \cmark & \xmark & \cmark & 1,510 & LEGAL-BERT~\cite{chalkidis2020legal} \\

    % 390+16
    \rowcolor{blue!15} 
    CC-BY-NC-ND-4.0\&3.0 & (\cmark) & \xmark & \xmark & \xmark & \xmark & \xmark & \xmark & \xmark & \xmark & \cmark & 406 & NonCommercial, NoDerivatives \\

    $^{\dagger}$CC0-1.0 & \cmark & \cmark & \textbf{!} & \cmark & \xmark & \xmark & \xmark & \xmark & \xmark & \xmark & 330 & BlueBERT~\cite{peng2019transfer} \\

    \rowcolor{blue!15} 
    C-UDA & \cmark & \cmark & \cmark & \xmark & \textbf{!} & \textbf{!} & \xmark & \xmark & \cmark & \cmark & 72 & Data for computational use only \\

    $^{\dagger}$PDDL & \cmark & \cmark & \cmark & \cmark & \xmark & \xmark & \xmark & \xmark & \xmark & \xmark & 50 & Database-specific license \\

    \rowcolor{blue!15}
    CC-BY-ND-4.0 & (\cmark) & \xmark & \xmark & \cmark & \xmark & \xmark & \cmark & \xmark & \xmark & \cmark & 47 & Disallow making derivatives \\

    *GFDL &  \multicolumn{10}{l|}{>Same as GPL, a free document license} & 30 & txtai-wikipedia \\

    \rowcolor{blue!15}
    *ODbL & \cmark & \cmark & \xmark & \cmark & \xmark & \xmark & \cmark & \cmark & \xmark & \cmark & 20 & Automatic relicensing \\

    *LGPL-LR & \cmark & \cmark & \xmark & \xmark & \textbf{!} & \textbf{!} & \cmark & \cmark & \xmark & \cmark & 19 & LGPL for linguistic resources \\ %如果仅仅发布embedding模型，那么属于“使用语言资源的作品”，如果包含了语言资源或者加密后的资源，那么属于“使用语言资源的衍生物”，包含在此license范围
    
    \rowcolor{blue!15} 
    ODC-By & \cmark & \cmark & \xmark & \cmark & \xmark & \xmark & \xmark & \xmark & \xmark & \cmark & 15 & Automatic relicensing \\
    %但是只对derivative dataset有定义，对于model来可能不适用

    %cdla-permissive-1.0 & \cmark & \cmark & \xmark & \cmark & \xmark & \xmark & \cmark & \cmark & \xmark & \cmark & 3 & Automatic relicensing \\

    %cdla-permissive-2.0\&1.0 & \cmark & \cmark & \xmark & \cmark & \xmark & \xmark & \cmark & \cmark & \xmark & \cmark & 5 & Automatic relicensing \\

    %tii-falcon-llm & \cmark & \cmark & \xmark & \cmark & \xmark & \xmark & \cmark & \cmark & \xmark & \cmark & 2 & Automatic relicensing \\

    %deepfloyd-if-license & \cmark & \cmark & \xmark & \cmark & \xmark & \xmark & \cmark & \cmark & \xmark & \cmark & 18 & Automatic relicensing \\

    \bottomrule
  \end{tabular}
  \vspace{-4mm}
\end{table*}

\subsubsection{Model Licensing Forms}
\label{sec:licensing}
As shown in TABLE~\ref{tab:licenses}, ML models are licensed in three main forms: as software (e.g., Apache, MIT, GPL), as a model (e.g., OpenRAIL), and as content/database (e.g., CC BY, PDDL).
The reason for the mixed use of licenses is the ambiguity in the dependency relationship between the ML code, model, and data.
Thinking in terms of software, ML models can be released with reproducable code and considered as a component of software.
So many Free and Open Source Software (FOSS) licenses~\cite{rosen2005open} are naturally deferred for licensing of models.
The most popular license is Apache-2.0, which is a permissive FOSS license that allows the freedom to make derivative works.
However, the model building process also relies on a massive amount of data~\cite{lecun2015deep} that may be licensed under different licenses, which can lead to license conflicts.
A real-world example is BERT~\cite{devlin2018bert}, which was published under the Apache-2.0 license but pre-trained on English Wikipedia documents that are licensed under CC BY-SA 3.0.
This changing of license violates the requirement of the CC BY-SA 3.0, which states that any contribution must be distributed under the \textbf{same license} as the original work.

%We don’t claim ownership of the content you create with GPT-2, so it is yours to do with as you please. We only ask that you use GPT-2 responsibly and clearly indicate your content was created using GPT-2.

Thinking in terms of content and database, some word embedding models like GloVe~\cite{pennington2014glove}, compute words representations based on licensed open linguistic resources.
These representations can be regarded as a translation of corpus and fall under the license of the original linguistic resources.
A more complex scenario arises when the model is fine-tuned with other data that has a different license, for example, fine-tune RoBERTa~\cite{liu2019roberta} (licensed under MIT) with SQuAD2~\cite{rajpurkar2016squad} (licensed under CC BY 4.0).
The tuned model can be interpreted as both derived works and combined works.

Not only limited to protecting the IP and controlling the diffusion of ideas, but AI companies and researchers are also concerned about licensees using their models for unethical purposes~\cite{jobin2019global, awad2018moral, yuste2017four}, which is usually not restricted by traditional licenses designed for software and content.
We can infer the concerns of unethical use of GPT-2~\cite{radford2019language} from its modified MIT license granted by its inventors, which states, \textit{We don't claim ownership of the content you create with GPT-2, so it is yours to do with as you please. We only ask that you use GPT-2 responsibly and clearly indicate your content was created using GPT-2.} 
However, such a statement lacks legal enforcement, and users may avoid accountability by convincing themselves that despite their efforts to minimize harm, they could not predict the AI artifact they generated would be used for harmful purposes.
On the other hand, the original licensing frameworks for software and content (e.g., MIT, CC BY) are not well suited to the data-driven ML. 
Many ML operations, such as training, fine-tuning, inference, and distillation, are not explicitly defined in these license clauses, leaving a potential legal loophole for licensees.

To address the unique challenges and considerations surrounding the use and distribution of ML models, several specific licenses for ML models have been proposed. 
CreativeML OpenRAIL-M license, proposed by Responsible AI~\cite{contractor2022behavioral}, is the most popular model-specific license on Hugging Face and enables legally enforceable responsible use.
By accepting this license, licensees must adhere to the use-based restrictions stated by the licensor, and these restrictions must also apply to derivative works.
With a multitude of different model licenses available, it becomes a challenging and tedious work to reuse them in bulk. 
It is, therefore, imperative to establish guidelines for selecting the licenses for models and other related components that are ready for query-based FL.

%The effect of copyleft-style behavioral-use clauses spreads the requirement from the original licensor on his/her wish and trust on the responsible use of the licensed artifact. This is why OpenRAILs require downstream adoption of the use-based restrictions by subsequent re-distribution and derivatives of the AI artifact, as a means to dissuade users of derivatives of the AI artifact from misusing the latter.

%OPT-175B/SEER LICENSE is not a copyleft license, as it does not require derivative works to be licensed under the same license or a compatible one. It is a proprietary license that allows users to use and reproduce the licensed models subject to certain restrictions.

\subsubsection{License Choosing}
\label{sec:choosing}
In query-based FL, the model community aims to promote the reuse of models contributed by users, which raises unique concerns about model licensing:
\begin{itemize}
  \item A model license ready for open FL paltforms should allow the \textbf{modification, combination and redistribution} of original works and any derived works.
  \item \textbf{Sublicensing} right should be granted to enable the republication of derived works after knowledge mining.
  \item Some licenses enforce the source of the derived works to be \textbf{disclosed} and prohibit their \textbf{commercial use}, which hinders model selling~\cite{chen2019towards}.
  \item Some licenses are \textbf{copyleft} (marked with * in TABLE~\ref{tab:licenses}), which means the derivatives must be licensed under the same license or a compatible license, leading to potential license conflicts and proliferation~\cite{gomulkiewicz2009open}.
  \item All granted rights are preferably \textbf{irrevocable} by the licensors~\cite{reddy2009jacobsen}.
\end{itemize}

Furthermore, it is important to consider the licensing of two other components when building and reusing models: data and algorithms, which may have entirely different license terms. 
Here, guided by the comparisons between different licenses outlined in TABLE~\ref{tab:licenses}, we can summarize the preferences for selecting licenses in query-based FL to minimize conflicts\textsuperscript{\ddag{III}}.
%Please see Appendix~III for our guidance on choosing licenses.
Once we obtain the right to relicense the modification models, the choice of a new license depends on the application scenario of models. 
We further provide a flowchart in Fig.~\ref{fig:flowchart}(a) to guide the license selection in the context of model query and model reusing.
In addition, to enhance the copyrightability of the reused models, we should avoid using any derivatives and generated content from models under proprietary licenses throughout the ML reusing lifecycle\textsuperscript{\ddag{III-D}}.

For now, we have provided a comprehensive perspective and suggestions regarding the regulations and legal issues related to batch model reusing with only one piece missing: the definition of terms and corresponding clauses for different reusing mechanisms in different licenses.
The terms definition for model reusing in different licenses is a novel and interesting issue that is rarely discussed. 
For example, interpreting model reusing as creating derivatives or combinations would involve different clauses in the licenses.
To provide a better understanding of these implications, let's first provide an overview of typical model reuse mechanisms.

\section{Batch Model Reuse Mechanisms}
\label{sec:taxonomy}
\begin{figure*}[t]
    \centering
    \includegraphics[width=\linewidth]{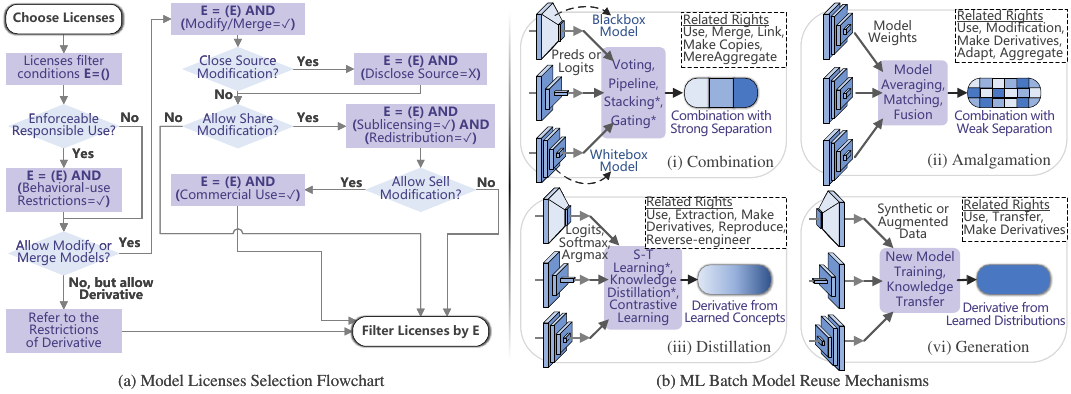}
    \vspace{-5mm}
    \caption{(a) Flowchart for model licenses selection in the context of model query and model reusing. (b) Proposed taxonomy categorizing batch model reuse mechanisms based on the reused results.}
    \label{fig:flowchart}
    \vspace{-4mm}
\end{figure*}

In this section, we delve into the typical batch model reuse mechanisms from a technical perspective and introduce a new taxonomy designed to address the mismatch between license terms and technical terms.
Therefore, instead of summarizing the batch model reuse mechanisms from a technological and algorithmic aspect, we propose grouping these mechanisms based on the classification of their resulting outputs for ease of justifying license clauses.
As shown in Fig.~\ref{fig:flowchart}(b), there are four categories of batch model reuse mechanisms: \textbf{Combination, Amalgamation, Distillation, and Generation}, each resulting in different forms of outputs and may correspond to different regulations in licenses.

\subsection{Combination}
\label{sec:combination}
Combination~\cite{zhou2012ensemble} is a straightforward way to reuse batch of models (base learners), in which multiple models jointly contribute to the output by combination strategies such as averaging, voting, stacking~\cite{jacobs1991adaptive, wolpert1992stacked}.
For regression estimates, averaging can improve the generalization by taking the mean of the outputs of all weak learners in a population. Additionally, the outputs of each learner can be weighted by extra parameters~\cite{perrone1995networks}, which can be determined by stacking estimators~\cite{wolpert1992stacked}, Bayesian approach~\cite{clarke2003comparing} or backpropagation of gating networks~\cite{jacobs1991adaptive}. 
% Bayesian Model Averaging (BMA) -> Stacking with probabilities
Voting is a workaround strategy for classification tasks and also applicable for stacking and gating.
Both stacking and gating rely on a holdout or validation dataset for calculating extra parameters, marked as * in Fig.~\ref{fig:flowchart}(b). 
The difference is that gating can adapt the weights of each model's estimation based on the inputs, providing better generalizability performance of the combined model.

There are many advantages of combination mechanisms from the perspective of FL.
First, the input spaces of base models can be unaligned, which is ideal for the scenario of vertical FL~\cite{wu2022practical} where each client may have inconsistent features in their data.
Secondly, especially for query-based FL, it can simultaneously support multiple types and heterogeneous models, which means that it does not rely on any prior assumptions of the models, such as whether they are DNNs or decision trees, released with raw weights (whitebox) or binary forms (blackbox).
Thirdly, the tasks of models can be different if we pipeline the base models end-to-end, which is usually overlooked as a combination mechanism of models. 
Pipelining can fully leverage the transferability of models to solve previously unexplored ML problems.
% 举例说明 pipeline
For instance, Gao \textit{et al.}~\cite{gao2022precise} proposed a zero-shot dense retrieval system named HyDE by pipelining a natural language generation (NLG) model~\cite{ouyang2022training} and a natural language understanding (NLU) model~\cite{izacard2022unsupervised}.
The generated content, which may lack factual grounding, from the NLG model is used as query embeddings to facilitate real document retrieval by the NLU model.
Similarly, through query-based FL, we can query a vicarious NLG model for a novel scenario, such as ProGen~\cite{madani2023large} for protein sequences generation, and quickly adapt this system to proteomics.
Not limited to that, we can query a batch of NLG models by a well-chosen filter condition and then combine models through averaging or gating to significantly expand the exploration space for knowledge discovery.

Lastly, the combinated models have strong separation from each other, meaning that we can add or remove a batch of models without significant changes to the remaining ones.
Meanwhile, combination mechanisms do not rely on the transparency of models and support blackbox sharing. 
Thus, the base model can establish loose connections with other models only through run scripts, providing revocability of such combination and circumvention of the restrictions of licenses.
On the other hand, instead of being treated as a challenge for FL~\cite{ma2022state}, the statistical heterogeneity and model heterogeneity nature of these crowdsourced models can actually enrich population diversity, which is crucial for creating a good ensemble~\cite{maclin1995combining, opitz1995generating}. 

However, the storage consumption of combined models increases linearly with the number of base learners, which can strain the communication resources of a collaborative model training network. 
As a result, alternative approaches involve amalgamating or merging multiple models to create a new consensus model. 
In the following, we provide a summary of these methods.

\subsection{Amalgamation}
\label{sec:amalgamation}
Amalgamation involves combining models through model parameters granularity operations, such as median~\cite{blanchard2017machine, pillutla2022robust} and coordinate-wise averaging with consideration of heterogeneity~\cite{mcmahan2017communication, li2018federated}, security~\cite{sun2019can}, scalability~\cite{reisizadeh2020fedpaq}, matching~\cite{wang2020federated, yu2021fed2}, specificality~\cite{gudur2020resource}, generalizability~\cite{qu2022generalized}, resulting in a combination with weak separation.
This reusing approach is widely used in FL works and is often referred to as "aggregation" procedures for local models.
%Here, we avoid using the term "aggregation" to distinguish it from "combination" given the latter is often used interchangeably with "combination" in software licenses (e.g. Artistic, GPL).
However, in order to avoid confusion with the term "combination," which is frequently used interchangeably with "aggregation" in software licenses (e.g., Artistic, GPL), we opt to use the term "amalgamation" instead.

FedAvg~\cite{mcmahan2017communication} is the vanilla model averaging method in FL with many follow-up works. 
For instance, Sun \textit{et al.}~\cite{sun2019can} introduced applying norm thresholding of local model updates to defend against backdoor attacks.
Similarly, Blanchard \textit{et al.}~\cite{blanchard2017machine} proposed using more robust median-based amalgamation for resilience against Byzantine behavior.
Consider the ordering of parameters, Wang \textit{et al.}~\cite{wang2020federated} match and average the NNs parameters layer-wise across clients, based on their similarities.
Yu \textit{et al.}~\cite{yu2021fed2} further attribute the misalignment in FL models to the non-IID training data, and propose allocating an independent structure for each class and updating models through a feature paired averaging strategy.

While amalgamation mechanisms can achieve a balance between model performance and resource efficiency by maintaining only one global model, they often rely on multiple rounds of communication to converge, which is not applicable in a query-based FL scenario.
Therefore, instead of trying to directly concatenate multiple sources of models while dealing with intricate parameter mismatch, transferring the latent knowledge learned by local models to a new model is a good alternative (i.e. Federated Distillation~\cite{jeong2018communication, jin2023feddyn}).

Another direction of model amalgamation is leveraging Bayesian nonparametrics to learn the shared global latent structures among local models~\cite{yurochkin2019bayesian, yurochkin2019statistical, lam2021model}. 
These methods, known as Model Fusion, can identify distributions of neural components across local models and only fuse the components with the same distribution, which can be regarded as a model compression between FedAvg (coordinate-wise averaging) and combination (w/o averaging).
However, the model fusion strategies rely on multiple communication rounds to boost the fusion efficiency, and the model performance of one-shot fusion is even worse than that of Ensemble.
Most recently, Su \textit{et al.}~\cite{su2023one}, inspired by null-space in continual learning~\cite{wang2021training, kong2022balancing}, propose MA-Echo which leverages layer-wise projection matrices to preserve the original loss of local models after amalgamation.
Their results present a moderate improvement in one-shot setting compared to FedAvg and ensemble.

Unfortunately, this improvement is not consistently observed in multiple-round experiments.
Meanwhile, to tackle the issue of catastrophic forgetting, FedPR~\cite{feng2023learning} follows similar ideas to facilitate the server's learning of visual prompts from clients for MRI reconstruction applications, but the improvement is limited even in multi-round setting. %compared to FedAvg

It is worth noting that our taxonomy is based on the form of the resulting model, which may not be entirely consistent with the terminology used in the technical perspective.
For example, Bayes Model Averaging (BMA)~\cite{clarke2003comparing} estimates posterior probabilities of each model given the observed data, which results in a separable weighted model. 
Therefore, it should be classified as Combination instead of Amalgamation like FedAvg.
This novel taxonomy method is useful for analyzing compatibility with licenses. 
For example, the coordinate-wise operations or the fusion of model parameters generate fine-grained combinations of models that are almost irreversible, which corresponds to clauses such as adapt, modify, dynamic link, and etc., in software licenses.

\begin{table*}[t]
  \centering
  \scriptsize
  \caption{Summary of privacy-preserving \textbf{Distillation} works in the field of FL. Some works are listed multiple times because they contain multiple KD procedures with different strategies. Works with naming conflicts are distinguished by subscript.}
  \label{tab:distillation}
  \begin{tabular}{|ll|p{12cm}|}
    \hline
    \multicolumn{2}{|l|}{Strategies} & FL Studies \\ \hline
    \multicolumn{1}{|l|}{\multirow{2}{*}{KD@Server}} & w/ Validation Set & FedED~\cite{sui2020feded} \\ \cline{2-3} 
    \multicolumn{1}{|l|}{} & w/ Unlabeled Data & FedDF~\cite{lin2020ensemble}, One-shot FL~\cite{guha2018one}, FedBE~\cite{chen2020fedbe}, PerAda~\cite{xie2023perada}, FedET~\cite{cho2022heterogeneous} \\ \hline
    \multicolumn{1}{|p{1.6cm}|}{\multirow{4}{*}{\shortstack{KD@Client \\ w/ Local Data}}} & from Global Model & FedFusion~\cite{yao2019towards}, FedKD\textsubscript{2}~\cite{wu2022communication}, MOON~\cite{li2021model}, FedNTD~\cite{lee2022preservation},  FedMLB~\cite{kim2022multi}, FedCAD~\cite{he2022class}, FedAlign\textsubscript{1}~\cite{mendieta2022local}, FedAlign\textsubscript{2}~\cite{zhang2023navigating}  \\ \cline{2-3} 
    \multicolumn{1}{|l|}{} & from Other Clients' Model  & FedMatch~\cite{jeong2020federated}, CCL~\cite{aketi2024cross}, FedProto\textsubscript{1}~\cite{michieli2021prototype}, FedProto\textsubscript{2}~\cite{tan2022fedproto} \\ \cline{2-3} 
    \multicolumn{1}{|l|}{} & from Self Model & FedFusion~\cite{yao2019towards}, FedDistill~\cite{jiang2020federated}, FedKD\textsubscript{2}~\cite{wu2022communication}, MOON~\cite{li2021model}, FCCL~\cite{huang2022learn}, pFedSD~\cite{jin2022personalized}, RSCFed~\cite{liang2022rscfed}, CCL~\cite{aketi2024cross} \\ \hline
    \multicolumn{2}{|l|}{KD w/ Public Unlabeled Datasets} & FedKT~\cite{li2021practical}, FedMD~\cite{li2019fedmd}, FedAD~\cite{gong2021ensemble}, FedMD-NFDP~\cite{sun2020federated}, FCCL~\cite{huang2022learn}, FedAUX~\cite{sattler2021fedaux}, RHFL~\cite{fang2022robust}, FedKD\textsubscript{1}~\cite{gong2022preserving}, Cronus~\cite{chang2021cronus}, KT-pFL~\cite{zhang2021parameterized}, DS-FL~\cite{itahara2021distillation} \\ \hline
    \multicolumn{2}{|l|}{KD w/ Generated Data (DFKD)} & DENSE~\cite{zhang2022dense}, FedCAVE-KD~\cite{heinbaugh2023data}, FedGen~\cite{zhu2021data},
    FedFTG~\cite{zhang2022fine} \\ \hline 
    \multicolumn{2}{|l|}{KD w/ Differential Privacy} & FedKC~\cite{wang2022fedkc}, FedSSL~\cite{fan2022private}  \\ \hline
    \end{tabular}
    \vspace{-5mm}
\end{table*}

\subsection{Distillation}
\label{sec:distillation}
Distillation was initially proposed by Hinton \textit{et al.}~\cite{hinton2015distilling} to transfer knowledge from a batch of independently trained neural network models (Specialists) to create a new Generalist model.
Their motivation was to explore the parallelization of training of specialists and improve the efficiency of distributed NNs modeling\cite{dean2012large}.
Each specialist only learns fine-grained distinctions of a subset of classes, which is very similar to the non-IID setting in FL~\cite{liqb2022federated}.
By using Knowledge Distillation (KD), we can also compress wide and deep teacher networks into lightweight student networks~\cite{romero2015fitnets}, which is promising for addressing system heterogeneity in cross-device FL~\cite{lim2020federated}.
Therefore, it is natural to extend the KD technologies to FL filed~\cite{wu2022communication}.
%~\cite{jiang2020federated, li2021practical, li2019fedmd, wu2022communication, chen2020fedbe, lin2020ensemble, gong2021ensemble, sun2020federated}.
Many recent FL works~\cite{aketi2024cross, luo2022fediris, tan2022fedproto}
%li2019fedmd, gong2021ensemble, sun2020federated, huang2022learn, fang2022robust, gong2022preserving, luo2022fediris, sui2020feded, chang2021cronus, he2020group, zhang2021parameterized, itahara2021distillation, aketi2024cross, michieli2021prototype, tan2022fedproto}
solely leverage KD without following the model averaging paradigm of FedAvg, we leave this discussion for later. %TODO

Despite directly retraining a Generalist model through KD, an alternative approach is to construct an ensemble of knowledge.
For example, Furlanello \textit{et al.}~\cite{furlanello2018born} consecutively generate student models with the guidance of knowledge distilled from earlier generations and find that the ensemble of multiple generations of internal models achieves state-of-the-art performance.
Dvornik \textit{et al.}~\cite{dvornik2019diversity} leverage the distilled knowledge from each learner to encourage cooperation and prediction diversity within the population, which leads to better ensemble results.
In the context of FL, the \emph{main advantage} of distillation is the decoupling between KD and knowledge learning. 
This allows us to split the model architecture for the purpose of system heterogeneity and efficiency~\cite{vepakomma2019split, thapa2022splitfed}.
Moreover, the well-learned knowledge from clients only needs to be communicated once~\cite{gong2021ensemble, gong2022preserving}, while the server can perform multiple epochs of local training to complete the transfer.

The drawback of KD is that it is data-dependent and the shared knowledge may be extracted from local sensitive data, which exposes a new attack surface for potential model inversion attacks~\cite{kim2020multiple, fredrikson2015model}.
To mitigate this issue, some efforts~\cite{wang2022fedkc, fan2022private} have been made to add differential privacy noise~\cite{dwork2006differential} to the shared content.

In general, there are three mechanisms for avoiding the sharing of sensitive knowledge.
First, push the KD procedure to the server-side, where the knowledge of local models is transferred to the global model through a validation set~\cite{sui2020feded} or unlabeled dataset~\cite{lin2020ensemble, guha2018one, chen2020fedbe, xie2023perada, cho2022heterogeneous} held by the server.
Second, we can keep the KD procedure at the client-side, allowing the knowledge of the global model~\cite{yao2019towards, wu2022communication,li2021model, lee2022preservation, mendieta2022local, kim2022multi, he2022class, zhang2023navigating}, other clients' models~\cite{jeong2020federated, aketi2024cross, michieli2021prototype, tan2022fedproto} or self-model~\cite{yao2019towards, jiang2020federated, wu2022communication, li2021model, huang2022learn, jin2022personalized, liang2022rscfed} to be transferred based on the local training data.
In the above two strategies, only model parameters are exchanged in the training network, which means they can provide the same level of privacy protection as traditional FL.

%The last mechanism assumes that a public unlabeled dataset is accessible to both the server and clients for KD~\cite{li2021practical, li2019fedmd, gong2021ensemble, sun2020federated, huang2022learn, sattler2021fedaux, fang2022robust, gong2022preserving, chang2021cronus, zhang2021parameterized, itahara2021distillation}.
The last mechanism is to assume that a public unlabeled dataset, which is non-sensitive, is accessible by both the server and clients for KD~\cite{li2021practical, li2019fedmd, gong2021ensemble, sun2020federated, huang2022learn, sattler2021fedaux, fang2022robust, gong2022preserving, chang2021cronus, zhang2021parameterized, itahara2021distillation}.
Sharing these extracted contents will not raise any privacy concerns, and only minimal communication is generated during KD for the purpose of aligning sample IDs.
In cases where such public datasets are not available on the server, a recent approach known as Data-Free Knowledge Distillation (DFKD)~\cite{lopes2017data} regenerates batches of data based on layer activation statistics or spectrum coefficients collected during training phase. 
Then, this synthetic data is used for distillation.
DENSE~\cite{zhang2022dense} is the first attempt to extend DFKD to FL. It leverages the ensemble of local models to guide the training of a data generator on the server, and the generated data is then used to distill the knowledge from local models to the global model.
FedCAVE-KD~\cite{heinbaugh2023data} leverages locally trained conditional autoencoders (CVAEs)~\cite{kingma2014auto} to generate samples based on the data distribution of clients. 
These CVAEs are sent to server used to construct a global generator via KD, which will later provide synthetic training data for the global discriminator.

It is worth noting that in the query-based FL setting, direct access to the original data is not available, thus the second mechanism mentioned earlier cannot be directly applied.
A circumvention method is to train a generator following the inspiration of DFKD. 
Fortunately, this is praticable if the workflow and history information of modeling are tracked and queryable, as we advocated in \S\ref{sec:how2query}.
Recalling that, as shown in Fig.~\ref{fig:flowchart}(b), Generation is the last category in our taxonomy.
Actually, such a hybrid model reuse strategy is quite common in FL. 
For example, the previously mentioned DENSE~\cite{zhang2022dense} incorporates three model reuse mechanisms: Combination (creating an ensemble), Generation (generating synthetic data), and Distillation.
Therefore, our taxonomy can cover traditional FL works, such as FedAvg and MOON~\cite{li2021model}, as well as the broad sense FL, including Federated Distillation~\cite{jeong2018communication, jin2023feddyn} and Ensemble Learning~\cite{shi2023fed, wang2023data}.
We provide a comparison of these hybrid works in \S\ref{sec:hybrid}. 
The summarization of above privacy-perserving KD works is given in TABLE~\ref{tab:distillation}.

% the model trained in one specific domain cannot well generalize to other domains
% KD 是用于加速网络训练，属于分布式机器学习范畴

% KD 的优点：不需要相同的模型结构，缺点：依赖数据, Data-dependent KD, extracted soft labels 参考：Related work of DYNAFED: Tackling Client Data Heterogeneity with Global Dynamics 

\subsection{Generation}
\label{sec:generation}
Generation is designed to generate synthetic samples that resemble the original data distribution by building a probabilistic model~\cite{geyer1992practical} or deep learning model~\cite{kingma2014auto, goodfellow2020generative, cao2022survey} that can capture the underlying distribution pattern and latent structure of original data.
Generally speaking, generation techniques can be classified into three categories: data-level, probabilistic, and representation-based approaches.
Data-level approaches involve sample granularity operations such as interpolation~\cite{chawla2002smote, zhangmixup} and augmentation~\cite{wong2016understanding} to generate synthetic features based on the original feature space of data and share.
Even though these methods are training-free and easy to implement, they cannot be directly applied to the FL setting due to privacy concerns.
Recent proposed FedMix~\cite{yoon2021fedmix} aims to alleviate the negative effect of non-IID data by using \textit{mixup}~\cite{zhangmixup}, where the average of local data is linearly interpolated with the training data to generate augmented samples.
However, the potential risk of data leakage when sharing the mixup data is not comprehensively evaluated in the original work.

Probabilistic approachs aim to estimate the real data distribution using probabilistic method.
For example, Markov Chain Monte Carlo (MCMC)~\cite{geyer1992practical} methods construct a Markov Chain that converges to the desired target distribution by iteratively proposing new states based on the current state of the chain and acceptance probabilities.
Gaussian Mixture Models (GMM) assumes the data is generated from a mixture of Gaussian distributions and can generate new samples by sampling from the learned distributions.
To generate the high-dimensional structured data, representation-based approaches try to reconstruct the data from laten feature space.
For instance, Variational Autoencoders (VAEs)~\cite{kingma2014auto} learn the distribution of the latent representation space given the observed data and then use a decoder network to reconstruct data based on sampled latent representations.
Generative Adversarial Networks (GANs)~\cite{goodfellow2020generative} train a generator network to produce samples that resemble realistic data by optimizing an adversarial objective against a discriminator network.

Compared to the other model reuse mechanisms, generation has three unique advantages.
The first advantage is visualization and verification. Unlike the extracted knowledge in distillation, the quality of generated content can be visualized and validated by humans. 
This capability aids in assessing the contributions made by participants in terms of generating valuable content.
Second, the flexibility of generation methods allow us to generate data with any desired amount or class, which enables more effective handling of imbalanced~\cite{chawla2002smote} and non-IID~\cite{zhang2022fine} data.
The third advantage is multi-format sharing. Participants have the freedom to choose the form of their contributions. For example, they can upload the learned generative models in source code (e.g., Stable Diffusion~\cite{rombach2022high}, GPT-2~\cite{radford2019language}) or binary form, upload synthetic data, or provide model inference APIs like ChatGPT.
This sharing policy can greatly empower the model community in open FL platforms, fostering collaboration and knowledge sharing.

Given the aforementioned advantages, generation methods have been extensively studied in the field of FL. 
As summaried in TABLE~\ref{tab:generation}, these works can be classified into two main categories: generation for training~\cite{zhang2021subgraph, cheng2023gfl, hao2021towards, cha2019federated, yu2023turning, heinbaugh2023data, yang2023exploring, liu2021feddg, pi2023dynafed, liz2022federated, diao2022semifl}, generation for KD~\cite{zhang2022dense, chen2020fedbe, zhu2021data, zhang2022fine, jeong2018communication, jin2023feddyn, heinbaugh2023data, zhang2022fedzkt, fan2022private}, and serving three purposes: enriching the training set~\cite{zhang2022dense, chen2020fedbe, zhang2021subgraph, cheng2023gfl, cha2019federated, jin2023feddyn, zhang2022fedzkt}, improving generalization ability~\cite{zhu2021data, zhang2022fine, hao2021towards, jeong2018communication, yu2023turning, heinbaugh2023data, liu2021feddg, pi2023dynafed, liz2022federated}, and enabling semi-supervised learning~\cite{yang2023exploring, diao2022semifl, fan2022private}.
As an example of generation for training, FedSage+~\cite{zhang2021subgraph} trains a missing neighbors generator to mend the links between cross-subgraph nodes, thereby increasing the connectivity of local data and benefiting from this collaboration across clients.
Previously mentioned FedCAVE-KD~\cite{heinbaugh2023data} is an example of generation for KD, where locally trained CVAEs and local label distributions are uploaded to the server for DFKD, ensuring privacy while also enhancing generalization of global model.
Another example is FedGen~\cite{zhu2021data}, where the generator is maintained by the server and sent to clients in each round.
The generator has knowledge about the global view of the data distribution, which is used to KD into the local models, thereby enhancing their generalizability.
The last application of generation is in semi-supervised learning, which is a common real-world scenario where the client data is unlabeled.
For example, FedDISC~\cite{yang2023exploring} leverages the average and cluster centroids of hidden representations across pseudo-labels as input to a pre-trained diffusion model, aiming to generate high-quality samples for training.

In fact, the three purposes of generation correspond to three types of data heterogeneity in FL~\cite{liqb2022federated}: Quantity Skew, Label Distribution Skew (non-IID), and Missing Labels, which are challenging to address with traditional model amalgamation methods.
The hybrid model reusing strategies, which leverage each other's strengths, have become a common paradigm in recent FL studies~\cite{cheng2023gfl, xie2023perada, yu2023turning, jin2023feddyn, heinbaugh2023data, yang2023exploring, zhang2023navigating}.
Therefore, in the next section, we will summarize the popular hybrid model reusing studies in FL and then filter the studies that are suitable for query-based FL platforms.

\begin{table}[t]
  \centering
  \tiny
  \caption{Classification of \textbf{Generation} works in the field of FL. Some of the works are also listed in TABLE~\ref{tab:distillation} because they utilize hybrid model reuse mechanisms.}
  \label{tab:generation}

  \begin{tabular}{|l|p{2.1cm}|p{2.3cm}|p{1.45cm}|}
    \hline
    & Enriching Training Set & Improving Generalization  & Enabling Semi-supervised Learning  \\ \hline
   
    \multicolumn{1}{|l|}{For Training} & FedSage+~\cite{zhang2021subgraph}, GFL~\cite{cheng2023gfl}, \newline FRD~\cite{cha2019federated} & Fed-ZDA~\cite{hao2021towards}, FedDG~\cite{liu2021feddg}, FOSTER~\cite{yu2023turning}, DynaFed~\cite{pi2023dynafed}, SDA-FL~\cite{liz2022federated}, FedMix~\cite{yoon2021fedmix}, \newline FedCAVE-Ens~\cite{heinbaugh2023data} & FedDISC~\cite{yang2023exploring}, \newline SemiFL~\cite{diao2022semifl}  \\ \hline

    \multicolumn{1}{|l|}{For KD} & DENSE~\cite{zhang2022dense}, FedBE~\cite{chen2020fedbe}, \newline FedDyn~\cite{jin2023feddyn}, FedZKT~\cite{zhang2022fedzkt} & FedGen~\cite{zhu2021data}, FedFTG~\cite{zhang2022fine}, FD+FAug~\cite{jeong2018communication}, FedCAVE-KD~\cite{heinbaugh2023data} & FedSSL~\cite{fan2022private} \\ \hline
  \end{tabular}
  \vspace{-5mm}
\end{table}

%The third advantage is free-sharing. As mentioned in Section~\ref{sec:generated content}, many deep generative models, such as Stable Diffusion~\cite{rombach2022high} and ChatGPT, do not claim ownership rights over the generated content. Furthermore, most OSS licenses do not have defined terms for computer-generated content. Therefore, it is legally permissible to freely reuse synthetic data.

% 三种方法：1，data level（如SMOTE, augmentation）2, 概率模型（MCMC, 自回归模型） 3，sample from latent representation（GAN，CVAE）
% 优点：1，生成的内容质量容易被校验（相比KD）2，可以掌控生成物的数量 3，弱连接(甚至不需要保留模型)
% 生成物的用法：1，重新训练 2，KD 3，提升模型鲁棒性和平衡性

%也可以是基于数据的生成

\subsection{Hybrid Model Reusing and Model Licenses}
\label{sec:hybrid}
Following the taxonomy we introduced in \S\ref{sec:taxonomy}, it can be observed that almost all FL studies can be regarded as a permutation of four model reuse mechanisms: Combination, Amalgamation, Distillation, and Generation.
To enhance our understanding of current model reusing studies and identify methods applicable for constructing an open FL platform, we provide a comprehensive summary in TABLE~\ref{tab:flreuse}.
We employ different colors and fonts in TABLE~\ref{tab:flreuse} to emphasize the distinctions among studies, while certain processes have been omitted without ambiguity.
In addition, we have listed the main goals of each study, and only those with explicit designs, experiments, or proof are counted.
Please refer to the table caption for the explanation of our denotations.
As an example, the process of RSCFed~\cite{liang2022rscfed} involves the following steps: 
\ding{172} KD, the knowledge is extracted as the softmax values from the self model based on local private data (ref. TABLE~\ref{tab:distillation}) and performed at the client-side;
\ding{173} Model exponential moving averaging performed at the client-side;
\ding{174} Simple Model Averaging performed two times at the server-side.
These processes are repeated across multiple communication rounds until completion.
In this way, we can categorize these works and make intuitive comparisons between them\textsuperscript{\ddag{IV-A}}.
%Due to page limits, we leave the detailed analysis in Appendix~IV.A.

Through disassembling hybrid model reusing into the four mechanisms, we can further analyze the corresponding clauses in model licenses perspective for FL studies.
Then we can easily identify the applicable clauses for different reusing mechanisms\textsuperscript{\ddag{IV-B}} and analyze potential licenses conflicts\footnote{\href{https://github.com/Xtra-Computing/ModelGo}{We developed ModelGo to implement this idea.}}.
%We present the detailed methodology in Appendix~IV.B.

\begin{table*}[htp]
  \centering
  \scriptsize
  \caption{Comparative analysis of FL studies categorized by our taxonomy for batch model reuse mechanisms.
  Studies \textcolor{red}{applicable} and \textcolor{orange}{conditional applicable} to query-based FL are marked with different colors;
  \textcolor{purple}{Purple} denotes operations completed on \textcolor{purple}{Server}, or knowledge distilled from \textcolor{purple}{Global or Consensus Model}; 
  \textcolor{blue}{Blue} denotes operations completed on \textcolor{blue}{Clients}, or knowledge distilled from \textcolor{blue}{Local, Personlized, or Generative Models}; 
  \textbf{Knowledge} or \textbf{Generated Content} based on \textbf{Public, Proxy or Generated} data, and \textit{Knowledge} or \textit{Generated Content} based on \textit{Local, Private or Sensitive} data;
  [~]*1: one round of communication (aka one-shot), [~]*N: multiple rounds of communication, processes ahead ...[~] are performed only once (i.e. preprocessing), processes inside [...] are main functional part; Slash "/": model training based on \textit{private} data, Comma ",": model training based on \textbf{non-sensitive} data; Goals of works: \textbf{E}fficiency \textbf{H}eterogeneity \textbf{P}rivacy.}
  %\vspace{-6mm}
  \label{tab:flreuse}
  \begin{tabular}{|p{2.05cm}|p{1.36cm}|p{1.56cm}|p{4.35cm}|p{2.77cm}|p{1.5cm}|p{0.35cm}|}
    \hline
    \rowcolor[gray]{.8}
    \multicolumn{1}{|c|}{FL Studies} & \multicolumn{1}{c|}{\textbf{C}ombination} & \multicolumn{1}{c|}{\textbf{A}malgamation} & \multicolumn{1}{c|}{\textbf{D}istillation} & \multicolumn{1}{c|}{\textbf{G}eneration} & \multicolumn{1}{c|}{Process}& \multicolumn{1}{c|}{Goals} \\ \hline
    FedAvg~\cite{mcmahan2017communication} & n/a &  \textcolor{purple}{Model Avg}  & n/a & n/a & [\textcolor{blue}{/}\textcolor{purple}{A}]*N & EH   \\ \hline

    \rowcolor[gray]{.9}
    \textcolor{red}{FedAD}~\cite{gong2021ensemble} & n/a & n/a & \textcolor{purple}{KD} \textcolor{blue}{\textbf{Attention, Logits}}  & n/a & \textcolor{blue}{/}[\textcolor{purple}{D}]*1 & HP \\ \hline %CC
  
    \textcolor{red}{FedKD\textsubscript{1}}~\cite{gong2022preserving} & n/a & n/a & \textcolor{purple}{KD} \textcolor{blue}{\textbf{Weighted Logits}} & n/a &\textcolor{blue}{/}[\textcolor{purple}{D}]*1 & EHP \\ \hline %CC

    \rowcolor[gray]{.9}
    \textcolor{red}{FedED}~\cite{sui2020feded} & n/a & n/a & \textcolor{purple}{KD} \textcolor{blue}{\textbf{Logits Avg on Global Validation Data}} & n/a &[\textcolor{blue}{,}\textcolor{purple}{D}]*N & EHP \\ \hline %CC 

    FedIris~\cite{luo2022fediris} & n/a & n/a & \textcolor{blue}{KD \textit{Hidden}} & n/a &[\textcolor{blue}{D}]*N & H \\ \hline %CC

    \rowcolor[gray]{.9}
    FedProto~\cite{tan2022fedproto} & n/a & n/a & \textcolor{blue}{KD \textit{Per-Class Hidden Avg}} & n/a &[\textcolor{blue}{/D}]*N & EH \\ \hline %CC

    FedMD~\cite{li2019fedmd} & n/a & n/a & \textcolor{blue}{KD \textbf{Logits Avg}} & n/a &[\textcolor{blue}{D/}]*N & H \\ \hline %CC

    \rowcolor[gray]{.9}
    FedMD-NFDP~\cite{sun2020federated}& n/a & n/a & \textcolor{blue}{KD \textbf{Logits/Softmax/Argmax Avg}} & n/a & \textcolor{blue}{/}[\textcolor{blue}{D/}]*N & HP \\ \hline %CC

    DS-FL~\cite{itahara2021distillation}& n/a & n/a & \textcolor{blue}{KD \textbf{Entropy Reduced Logits Avg}} & n/a &[\textcolor{blue}{/D}]*N & EH \\ \hline %CC

    \rowcolor[gray]{.9}
    \textcolor{red}{RHFL}~\cite{fang2022robust} & n/a & n/a & \textcolor{blue}{KD \textbf{Weighted Logits}} & n/a & \textcolor{blue}{/}[\textcolor{blue}{D}]*N & EH \\ \hline

    KT-pFL~\cite{zhang2021parameterized} & n/a & n/a & \textcolor{blue}{KD \textbf{Learned Weighted Softmax}} & n/a &[\textcolor{blue}{/D}]*N & EH \\ \hline

    \rowcolor[gray]{.9}
    Cronus~\cite{chang2021cronus} & n/a & n/a & \textcolor{blue}{KD} \textcolor{blue}{\textbf{Robust Mean Estimation of Softmax}} & n/a & \textcolor{blue}{/}[\textcolor{blue}{/D}]*N & EHP \\ \hline
    
    FedDISC~\cite{yang2023exploring} & n/a & n/a & n/a & \textcolor{purple}{\textit{Synthetic Data}} &[\textcolor{purple}{G}]*1\textcolor{purple}{,} & EH \\ \hline

    \rowcolor[gray]{.9}
    FRD~\cite{cha2019federated} & n/a & n/a & n/a & \textcolor{purple}{\textit{Mixup Data}} &[\textcolor{purple}{G}]*1\textcolor{purple}{,} & E \\ \hline

    \textcolor{red}{One-Shot FL}~\cite{guha2018one} & \textcolor{purple}{Output Avg} & n/a & \textcolor{purple}{KD} \textcolor{blue}{\textbf{Softmax}} & n/a & \textcolor{blue}{/}[\textcolor{purple}{CD}]*1 & EP \\ \hline %CC

    \rowcolor[gray]{.9}
    FedDF~\cite{lin2020ensemble} & n/a & \textcolor{purple}{Model Avg} & \textcolor{purple}{KD}  \textcolor{blue}{\textbf{Logits Avg}} & n/a & [\textcolor{blue}{/}\textcolor{purple}{AD}]*N & HP \\ \hline %CC

    PerAda~\cite{xie2023perada} & n/a & \textcolor{purple}{Adapter Avg} & \textcolor{purple}{KD} \textcolor{blue}{\textbf{Logits Avg}} & n/a & [\textcolor{blue}{//}\textcolor{purple}{AD}]*N & EHP \\ \hline

    \rowcolor[gray]{.9}
    FedFiMa~\cite{che2022federated} & n/a & \textcolor{purple}{Rep. Layer Avg} & \textcolor{blue}{KD} \textcolor{blue}{\textbf{Hidden Avg}} & n/a & [\textcolor{blue}{/}\textcolor{purple}{A}\textcolor{blue}{D}]*N & EH \\ \hline

    FedFusion~\cite{yao2019towards} & n/a & \textcolor{purple}{Model Avg} & \textcolor{blue}{KD} \textcolor{blue}{\textit{Hidden}}, \textcolor{purple}{\textit{Hidden}} & n/a & [\textcolor{blue}{D}\textcolor{purple}{A}]*N & E \\ \hline %CC

    \rowcolor[gray]{.9}
    FedNTD~\cite{lee2022preservation} & n/a & \textcolor{purple}{Model Avg} & \textcolor{blue}{KD} \textcolor{purple}{\textit{Not-True Classes Softmax}} & n/a & [\textcolor{blue}{D}\textcolor{purple}{A}]*N & H \\ \hline %CC

    FedKC~\cite{wang2022fedkc} & n/a & \textcolor{purple}{Model Avg} & \textcolor{blue}{KD} \textcolor{blue}{\textit{Clustered Hidden Avg}} & n/a & [\textcolor{blue}{D}\textcolor{purple}{A}]*N & HP \\ \hline %CC

    \rowcolor[gray]{.9}
    FedDistill~\cite{jiang2020federated} & n/a & \textcolor{purple}{Model Avg} & \textcolor{blue}{KD \textit{Softmax of Latest Local Model}} & n/a &[\textcolor{blue}{/D}\textcolor{purple}{A}]*N & H \\ \hline %CC

    pFedSD~\cite{jin2022personalized} & n/a & \textcolor{purple}{Model Avg} & \textcolor{blue}{KD \textit{Softmax of Pervious Local Model}} & n/a &[\textcolor{blue}{/D}\textcolor{purple}{A}]*N & H \\ \hline

    \rowcolor[gray]{.9}
    FedMLB~\cite{kim2022multi} & n/a & \textcolor{purple}{Model Avg} &\textcolor{blue}{KD} \textcolor{purple}{\textit{Softmax, Scaled Softmax}} & n/a & [\textcolor{blue}{/D}\textcolor{purple}{A}]*N & EH \\ \hline %CC

    FedAlign\textsubscript{1}~\cite{mendieta2022local} & n/a &\textcolor{purple}{Model Avg}&\textcolor{blue}{KD}  \textcolor{purple}{\textit{Lipschitz Constants}}~\cite{shang2021lipschitz} & n/a & [\textcolor{blue}{/D}\textcolor{purple}{A}]*N & EH \\ \hline %CC
    
    \rowcolor[gray]{.9}
    MOON~\cite{li2021model} & n/a & \textcolor{purple}{Model Avg} & \textcolor{blue}{Contrastive Learning} \textcolor{purple}{\textit{Hidden}}, \textcolor{blue}{\textit{Hidden}} & n/a &[\textcolor{blue}{/D}\textcolor{purple}{A}]*N & EH \\ \hline %CC

    FedCAD~\cite{he2022class} & n/a & \textcolor{purple}{Model Avg} & \textcolor{blue}{KD} \textcolor{purple}{\textit{Class-Wise Softmax}} & n/a & [\textcolor{blue}{/D}\textcolor{purple}{A}]*N & H \\ \hline %CC

    \rowcolor[gray]{.9}
    FedGKT~\cite{he2020group} & n/a & n/a & \textcolor{blue}{KD} \textcolor{purple}{\textit{Logits}} \textcolor{purple}{KD} \textcolor{blue}{\textit{Logits, Hidden, Argmax}} & n/a & [\textcolor{blue}{/D}\textcolor{purple}{/D}]*N & EH \\ \hline

    FCCL~\cite{huang2022learn} & n/a & n/a & \textcolor{purple}{Contrastive Learning} \textcolor{blue}{\textbf{Logits Avg}} \newline \textcolor{blue}{Continual Learning} \textcolor{blue}{\textit{Logits}} & n/a & [\textcolor{purple}{D}\textcolor{blue}{/D}]*N & H \\ \hline %CC

    \rowcolor[gray]{.9}
    \textcolor{orange}{GFL}~\cite{cheng2023gfl} & n/a & \textcolor{purple}{Model Avg} & n/a & \textcolor{blue}{\textbf{Synthetic Data}} & \textcolor{blue}{/G}[\textcolor{blue}{,}\textcolor{purple}{A,}]*N & HP \\ \hline

    FOSTER~\cite{yu2023turning} & n/a & \textcolor{purple}{Model Avg} & n/a & \textcolor{blue}{\textbf{Synthetic Outliers}} & \textcolor{purple}{,}[\textcolor{blue}{G/}\textcolor{purple}{A}]*N & EH \\ \hline

    \rowcolor[gray]{.9}
    FedDG~\cite{liu2021feddg} & n/a & \textcolor{purple}{Model Avg} & n/a & \textcolor{blue}{\textbf{Interpolated Data}} & [\textcolor{blue}{G/}\textcolor{purple}{A}]*N & H \\ \hline  % 基于振幅的插值，可以认为不涉及隐私数据

    SemiFL~\cite{diao2022semifl} & n/a & \textcolor{purple}{Model Avg} & n/a & \textcolor{blue}{\textit{Augmented and Mixup Data}} & [\textcolor{purple}{,}\textcolor{blue}{G/}\textcolor{purple}{A}]*N & H \\ \hline 

    \rowcolor[gray]{.9}
    NeighGen~\cite{zhang2021subgraph} \newline FedSage~\cite{zhang2021subgraph} & & \textcolor{purple}{Gradient Avg}\newline\textcolor{purple}{Model Avg} & n/a &\textcolor{blue}{\textit{Synthetic Node}} & [\textcolor{blue}{G/}\textcolor{purple}{A}]*N \newline [\textcolor{blue}{/}\textcolor{purple}{A}]*N & H \\ \hline

    Fed-ZDAC~\cite{hao2021towards} & n/a & \textcolor{purple}{Model Avg} & n/a & \textcolor{blue}{\textbf{Zero-shot Synthetic Data}} & [\textcolor{blue}{G/}\textcolor{purple}{A}]*N & HP \\
    Fed-ZDAS~\cite{hao2021towards} & n/a & n/a & n/a & \textcolor{purple}{\textbf{Zero-shot Synthetic Data}} & [\textcolor{blue}{/}\textcolor{purple}{G,A}]*N & \\ \hline

    \rowcolor[gray]{.9}
    FedMix~\cite{yoon2021fedmix} & n/a & \textcolor{purple}{Model Avg} & n/a & \textcolor{purple}{\textit{Mixup Data}} \textcolor{blue}{\textit{Mixup Data}} & [\textcolor{purple}{GA\textcolor{blue}{,}}]*N\textcolor{blue}{G}\textcolor{blue}{,} & HP \\ \hline

    FedDyn~\cite{jin2023feddyn} & n/a & n/a & \textcolor{purple}{KD} \textcolor{blue}{\textit{Hidden, Logits Avg}} & \textcolor{blue}{\textbf{Synthetic Data}} & [\textcolor{blue}{/G}\textcolor{purple}{D}]*N & E \\ \hline

    \rowcolor[gray]{.9}
    FD+FAug~\cite{jeong2018communication} & n/a & n/a & \textcolor{blue}{KD} \textcolor{blue}{\textit{Per-Class Logits Avg}} & \textcolor{blue}{\textbf{Synthetic Data}} & [\textcolor{purple}{/}\textcolor{blue}{G}\textcolor{blue}{/D}]*N & EP \\ \hline

    \textcolor{red}{FedCAVE-Ens}~\cite{heinbaugh2023data} & \textcolor{purple}{Collection}  & n/a & n/a & \textcolor{purple}{\textbf{Synthetic Data}} & \textcolor{blue}{/}\textcolor{purple}{C}[\textcolor{purple}{G,}]*1 & H \\
    \textcolor{red}{FedCAVE-KD}~\cite{heinbaugh2023data} & n/a & n/a & \textcolor{purple}{KD} \textcolor{blue}{\textbf{Softmax}} & & \textcolor{blue}{/}\textcolor{purple}{C}[\textcolor{purple}{GDG,}]*1 & H \\ \hline

    \rowcolor[gray]{.9}
    Fed-ET~\cite{cho2022heterogeneous} & n/a & \textcolor{purple}{Rep. Layer Avg} \newline \textcolor{purple}{Model Avg} & \textcolor{purple}{KD} \textcolor{blue}{\textbf{Argmax, Weighted Logits}} & n/a & [\textcolor{blue}{/}\textcolor{purple}{ADA}]*N & EH \\ \hline

    \textcolor{orange}{FedBE}~\cite{chen2020fedbe} & n/a & \textcolor{purple}{Model Avg} & \textcolor{purple}{KD} \textcolor{blue}{\textbf{Softmax Avg}} & \textcolor{purple}{\textbf{Synthetic Model}} & [\textcolor{blue}{/}\textcolor{purple}{AGD}]*1 & H \\ \hline %CC

    \rowcolor[gray]{.9}
    DynaFed~\cite{pi2023dynafed} & n/a & \textcolor{purple}{Model Avg} & n/a & \textcolor{purple}{\textbf{Synthetic Data}} & [\textcolor{blue}{/}\textcolor{purple}{A}]*N\textcolor{purple}{G}[\textcolor{purple}{A,}]*N & EH \\ \hline

    \textcolor{orange}{FedAUX}~\cite{sattler2021fedaux} & n/a & \textcolor{purple}{Model Avg} & \textcolor{blue}{Contrastive Learning} \textcolor{purple}{\textbf{Hidden}, \textit{Hidden}} \newline \textcolor{purple}{KD} \textcolor{blue}{\textbf{Weighted Logits}} & n/a & \textcolor{purple}{,}\textcolor{blue}{D}[\textcolor{purple}{A\textcolor{blue}{/}D}]*N \newline \textcolor{purple}{,}\textcolor{blue}{D}[\textcolor{purple}{A\textcolor{blue}{/}D}]*1 & HP \\ \hline %CC
    
    \rowcolor[gray]{.9}
    FedKD\textsubscript{2}~\cite{wu2022communication} & n/a & \textcolor{purple}{Gradient Avg} & \textcolor{blue}{KD} \textcolor{purple}{\textit{Hidden, Attention, Logits}} \newline \textcolor{blue}{KD} \textcolor{blue}{\textit{Hidden, Attention, Logits}} & n/a & [\textcolor{blue}{/D}\textcolor{purple}{A}\textcolor{blue}{D}]*N & EH \\ \hline %CC

    FedGen~\cite{zhu2021data} & n/a & \textcolor{purple}{Model Avg} & \textcolor{purple}{KD} \textcolor{blue}{\textbf{Softmax}} \textcolor{purple}{KD} \textcolor{blue}{\textbf{Logits Avg}} & \textcolor{blue}{\textbf{Augmented Data}} & \textcolor{purple}{D}[\textcolor{blue}{/G,}\textcolor{purple}{AD}]*N & EHP \\ \hline %CC

    \rowcolor[gray]{.9}
    SDA-FL~\cite{liz2022federated} & n/a & \textcolor{purple}{Model Avg} & \textcolor{purple}{KD} \textcolor{blue}{\textbf{Argmax}} & \textcolor{blue}{\textbf{Synthetic Data}}\newline \textcolor{blue}{\textit{Augmented Data}} & \textcolor{blue}{/G}[\textcolor{blue}{G/}\textcolor{purple}{AD}]*N & HP \\ \hline

    FedKT~\cite{li2021practical} & \textcolor{blue}{Voting} \textcolor{purple}{Voting} & n/a & \textcolor{blue}{KD} \textcolor{blue}{\textbf{Argmax}} \textcolor{purple}{KD} \textcolor{blue}{\textbf{Argmax}} & n/a & [\textcolor{blue}{/CD}\textcolor{purple}{CD}]*1 & HP \\ \hline %CC %是否属于KD还是train取决于数据是否由其他模型生成的

    \rowcolor[gray]{.9}
    FedMatch~\cite{jeong2020federated} & \textcolor{blue}{Voting} & \textcolor{purple}{Model Avg} & \textcolor{blue}{KD} \textcolor{blue}{\textit{Argmax}} & n/a & [\textcolor{blue}{C/D}\textcolor{purple}{AA}]*N \newline [\textcolor{purple}{,}\textcolor{blue}{CD}\textcolor{purple}{A}]*N & EH \\ \hline %CC

    FedFTG~\cite{zhang2022fine} & n/a & \textcolor{purple}{Model Avg} &\textcolor{purple}{KD} \textcolor{blue}{\textbf{Softmax}}, \textcolor{purple}{\textbf{Softmax}} \textcolor{purple}{KD} \textcolor{blue}{\textbf{Softmax}} & \textcolor{purple}{\textbf{Synthetic Data}} & [\textcolor{blue}{/}\textcolor{purple}{AGDD}]*N & EH \\ \hline

    \rowcolor[gray]{.9}
    RSCFed~\cite{liang2022rscfed} \newline (Unlabeled Case) & n/a & \textcolor{blue}{Model EMA}\newline \textcolor{purple}{Model Avg} &\textcolor{blue}{KD} \textcolor{blue}{\textit{Softmax}} & n/a & [\textcolor{blue}{DA}\textcolor{purple}{AA}]*N & H \\ \hline

    FedAlign\textsubscript{2}~\cite{zhang2023navigating} & n/a &\textcolor{purple}{Model Avg}&\textcolor{blue}{Contrastive Learning}  \textcolor{purple}{\textit{Argmax, Argmin}} & n/a & [\textcolor{blue}{DD}\textcolor{purple}{AA}]*N & EH \\ \hline

    \rowcolor[gray]{.9}
    FedSSL~\cite{fan2022private} & n/a & n/a & \textcolor{blue}{KD} \textcolor{blue}{\textbf{Softmax}} \newline \textcolor{blue}{KD} \textcolor{blue}{\textbf{Softmax}, \textit{Interpolated Softmax}} & \textcolor{blue}{\textbf{Synthetic Data}} & [\textcolor{blue}{/GD/GD}]*N & HP \\ \hline
    \textcolor{red}{DENSE}~\cite{zhang2022dense} & \textcolor{purple}{Collection} & n/a & \textcolor{purple}{KD} \textcolor{blue}{\textbf{Logits Avg}, \textit{Batch-Wise Statistics}}, \textcolor{purple}{\textbf{Softmax}} \textcolor{purple}{KD} \textcolor{blue}{\textbf{Softmax}}, \textcolor{purple}{\textbf{Softmax}} & \textcolor{purple}{\textbf{Synthetic Data}} & \textcolor{blue}{/}\textcolor{purple}{C}[\textcolor{purple}{G,DGD}]*1 & HP \\ \hline %CC

    \rowcolor[gray]{.9}
    FedZKT~\cite{zhang2022fedzkt} & n/a & n/a &\textcolor{purple}{KD} \textcolor{purple}{\textbf{Softmax}}, \textcolor{blue}{\textbf{Softmax Avg}} \newline \textcolor{purple}{KD} \textcolor{blue}{\textbf{Softmax Avg}} \textcolor{purple}{KD} \textcolor{purple}{\textbf{Softmax}} & \textcolor{purple}{\textbf{Synthetic Data}} & [\textcolor{blue}{/}\textcolor{purple}{GDGDGD}]*N & H \\ \hline

  \end{tabular}
\end{table*}

\section{Remaining Topics in Query-based FL}
\label{sec:remaining_qbfl}
\subsection{Model Protection}
\label{sec:ip_protect}
In the previous sections, we have extensively discussed the construction of a query-based FL platform from both technical and legal perspectives. 
However, model management and protection continues to be a significant concern, raising questions such as \textit{How can I protect my models from plagiarism after they are released?}
Our conclusion is using \textbf{dynamic fingerprinting strategies with blackbox verification support}\textsuperscript{\ddag{V}}, such as DeepJudge~\cite{chen2022copy} and Zest~\cite{jia2022zest}.% See Appendix V for the complete discussion.

\subsection{Limitations of Query-based FL}
\label{sec:limitations_qbfl}
In query-based FL systems, the processes of model production and reuse are decoupled to maximize the autonomy of each participant. 
However, this losse cooperation paradigm is no longer compatiable with online collaboration ML frameworks, which means that it cannot fully harness participants' communicational and computational resource to enhace the training performance.
Therefore, even though the server-dominated cooperation frameworks like FedAvg have limitations, as described in \S\ref{sec:limitations_FL}, it is still worthwhile to provide support for its underlying distributed ML training methods to enhance the flexibility and compatibility of our open FL platforms.
In the next section, we illustrate another cooperation framework named contract-based FL, which follows a mutual choice design philosophy similar to crowdsourcing platforms~\cite{vaughan2018making}.
It can serve as an extension of traditional FL and query-based FL.

\section{Contract-based Federated Learning}
\label{sec:contract}
\subsection{Overiew}
\label{sec:contract_overview}
\begin{figure}[t]
    \centering
    \includegraphics[width=\linewidth]{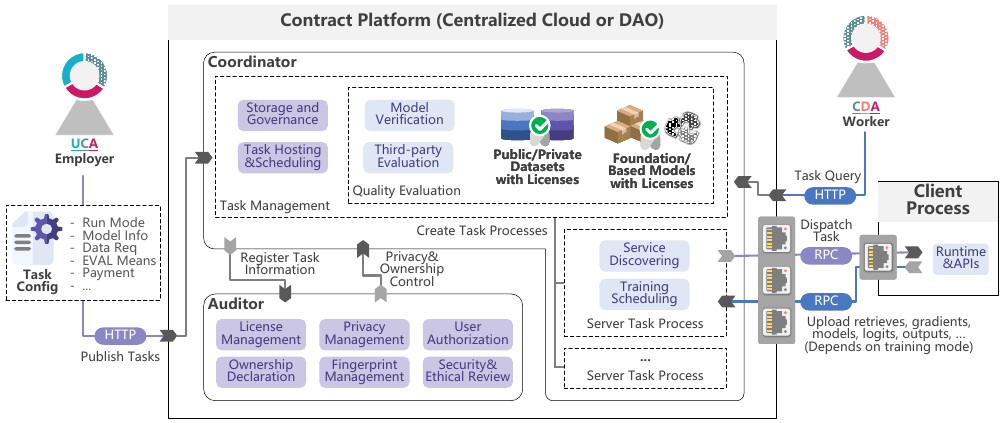}
    \caption{An overview of contract-based FL systems. (U: model User, C: Coordinator, D: Data owner, A: Auditor)}
    \label{fig:contract}
    \vspace{-5mm}
\end{figure}

Let us consider another open FL platform based on a contract-based cooperation framework, where cooperation can be built by publishing ML tasks and accepting ML collaboration requests.
An overview of this platform is presented in Fig.~\ref{fig:contract}, the major difference between traditional FL (ref Fig.~\ref{fig:fl}) and contract-based FL is that we involve a trustworthy third-party platform called Contract Platform to host and coordinate ML tasks for platform users.
Rather than directly pushing ML tasks from servers to clients, employers need to publish the tasks to the contract platform and wait for acceptance from workers, which involves a mutual choice procedure. 
The conditions and payment for participation are also appliable, and the final model will be audited and evaluated by the platform for fairness.
In addition, privacy-enhanced technologies~\cite{hesamifard2018privacy} and fingerprint management~\cite{chen2022copy} can also be implemented by the platform to protect the IP of platform users.
The properties of contract-based FL are: 1) \textbf{Opt-in}, as workers reserve the right to join or quit from training networks; 2) \textbf{Contractual}, enabling employers to define payment, organization mode, model quality criteria, rehire rules, etc. through contracts; 3) \textbf{Market-based}, where contracts are open and task pricing is influenced and determined by the market.

In short, contract-based FL retains all functionalities related to model training at the platform and only provides task publishing services to employers, who play the role of model training in traditional FL.
Furthermore, similar to model-reusing mechanisms (ref. \S\ref{sec:taxonomy}) that can be applied to query-based FL, our contract-based FL is not limited to an amalgamation-based collaborative training paradigm like FedAvg.
In fact, the contract platform is designed as a crowdsourcing platform among data miners for ML cooperation tasks, including distributed training, fine-tuning, and ensembling based on scratch or Foundation Models (FMs)~\cite{yuan2022decentralized}.
However, most current crowdsourcing platforms like Amazon Mechanical Turk and Appen\footnote{AMT: \url{https://www.mturk.com/}; Appen: \url{https://appen.com/crowd-2/}} primarily deal with human intelligence tasks such as data collection and annotation (aka. microtasks) rather than ML modeling tasks. 
It remains unexplored how to design and publish an ML task to seek crowd labor.
On the other hand, many studies~\cite{dias2022blocklearning, blythman2022decentralized, deng2021flex, guo2023blockchain, batool2022fl} and products~\cite{steeves2022incentivizing, ziller2021pysyft, mcconaghy2022ocean} emphasize monetizing ML activities through blockchain-based techniques and Decentralized Autonomous Organizations (DAOs). 
However, both monetization and decentralization are non-essential functions for our contract platforms\textsuperscript{\ddag{VI}}.
%, as we summarize and compare these works in Appendix~VI.

\begin{table*}[t]
    \centering
    \footnotesize
    \caption{Comparisons of typical model training frameworks with the intervention of a trustworthy platform. \colorbox{Employer!30}{\textasteriskcentered}, \colorbox{Platform!30}{\textasteriskcentered}, \colorbox{Worker!30}{\textasteriskcentered} indicate resources from \colorbox{Employer!30}{Employer}, \colorbox{Platform!30}{Platform}, and \colorbox{Worker!30}{Worker}, respectively. Resources in different color columns (e.g., \colorbox{Worker!30}{Data} in column \colorbox{Platform!30}{Platform}) represent transmission. $\bigstar$: Location of model training. \colorbox{Worker!30}{\textit{Resources}} received from other workers.}
    \vspace{-2mm}
    \label{tab:train}
    \begin{tabular}{|l|p{2.6cm}|p{4cm}|p{5cm}|}
    \hline
    & \multicolumn{1}{c|}{\cellcolor{Employer!30}{Employer}} & \multicolumn{1}{c|}{\cellcolor{Platform!30}{Platform}} & \multicolumn{1}{c|}{\cellcolor{Worker!30}{Worker}} \\ \hline
    
    Centralized ML & \colorbox{Employer!30}{Model} & \colorbox{Employer!30}{Model}\colorbox{Worker!30}{Data} $\bigstar$ & \colorbox{Worker!30}{Data} \\ \hline %Scratch

    FedAvg~\cite{mcmahan2017communication} & \colorbox{Employer!30}{Model} & \colorbox{Employer!30}{Model}\colorbox{Worker!30}{Gradient} & \colorbox{Employer!30}{Model}\colorbox{Worker!30}{Data} $\bigstar$ \\ \hline %Scratch

    DMoE~\cite{ryabinin2020towards} & \colorbox{Employer!30}{Gating NN}\colorbox{Employer!30}{Data} & \colorbox{Employer!30}{Gating NN}\colorbox{Employer!30}{Data}\colorbox{Worker!30}{Output} $\bigstar$ & \colorbox{Worker!30}{Model}\colorbox{Employer!30}{Data} $\bigstar$ \\ \hline %Scratch

    DeDES~\cite{wang2023data} & \colorbox{Worker!30}{Model Subset} & \colorbox{Worker!30}{Model} & \colorbox{Worker!30}{Model}\colorbox{Worker!30}{Data} $\bigstar$ \\ \hline %Scratch
    
    Borzunov \textit{et al.}~\cite{borzunov2022training} & \colorbox{Employer!30}{Model} & \colorbox{Employer!30}{Model}\colorbox{Platform!30}{Data}\colorbox{Worker!30}{Gradient} & \colorbox{Employer!30}{Model}\colorbox{Platform!30}{Data-Stream} $\bigstar$ \\ \hline %Scratch, DHT

    Moshpit SGD~\cite{ryabinin2021moshpit} & \colorbox{Employer!30}{Model}\colorbox{Employer!30}{Data} & \colorbox{Employer!30}{Model}\colorbox{Employer!30}{Data}\colorbox{Worker!30}{Gradient} & \colorbox{Employer!30}{Model}\colorbox{Employer!30}{Data-Loader} $\bigstar$ \\ \hline 

    VC-ASGD~\cite{atre2021distributed} & \colorbox{Employer!30}{Model}\colorbox{Employer!30}{Data} & \colorbox{Employer!30}{Model}\colorbox{Employer!30}{Data}\colorbox{Worker!30}{Gradient} & \colorbox{Employer!30}{Sub-model}\colorbox{Employer!30}{H-Data} $\bigstar$ \\ \hline

    SplitNN~\cite{vepakomma2019split} & \colorbox{Employer!30}{Model}\colorbox{Employer!30}{ID-Label} & \colorbox{Employer!30}{Model}\colorbox{Employer!30}{ID-Label}\colorbox{Worker!30}{Hidden} $\bigstar$ & \colorbox{Employer!30}{Sub-model}\colorbox{Worker!30}{V-Data}\colorbox{Platform!30}{Gradient} $\bigstar$ \\ \hline
    
    DT-FM~\cite{yuan2022decentralized} & \colorbox{Employer!30}{FM} & \colorbox{Employer!30}{FM}\colorbox{Platform!30}{Data} & \colorbox{Employer!30}{Sub-FM}\colorbox{Platform!30}{Micro-Batch}\colorbox{Worker!30}{\textit{Gradient}} $\bigstar$ \\ \hline

    FS-LLM~\cite{kuang2023federatedscope} & \colorbox{Employer!30}{FM} & \colorbox{Employer!30}{FM}\colorbox{Worker!30}{Adapter} & \colorbox{Employer!30}{FM}\colorbox{Worker!30}{Data} $\bigstar$ \\ \hline

    FedKSeed~\cite{qin2023federated} & \colorbox{Employer!30}{FM} & \colorbox{Employer!30}{FM}\colorbox{Platform!30}{Seed}\colorbox{Worker!30}{Scaler Gradient} & \colorbox{Employer!30}{FM}\colorbox{Platform!30}{Seed}\colorbox{Worker!30}{Data} $\bigstar$ \\ \hline

    Berdoz \textit{et al.}~\cite{berdoz2022scalable} & \colorbox{Employer!30}{Class} & \colorbox{Employer!30}{Class}\colorbox{Platform!30}{NN Structure}\colorbox{Worker!30}{Hidden} & \colorbox{Platform!30}{NN Structure}\colorbox{Worker!30}{Data}\colorbox{Platform!30}{Hidden AVG} $\bigstar$ \\ \hline

    CCL~\cite{aketi2024cross} & \colorbox{Employer!30}{Class} & \colorbox{Employer!30}{Class} & \colorbox{Worker!30}{Data}\colorbox{Worker!30}{\textit{Model}}\colorbox{Worker!30}{\textit{Hidden AVG}} $\bigstar$ \\ \hline

    FedProto~\cite{tan2022fedproto, michieli2021prototype} & \colorbox{Employer!30}{Class} & \colorbox{Employer!30}{Class}\colorbox{Worker!30}{Hidden} & \colorbox{Worker!30}{Data}\colorbox{Worker!30}{Model}\colorbox{Platform!30}{Hidden AVG} $\bigstar$ \\ \hline   

    Ocean~\cite{mcconaghy2022ocean} & \colorbox{Employer!30}{Model}\colorbox{Worker!30}{Output} & \colorbox{Worker!30}{Metadata} & \colorbox{Worker!30}{Metadata}\colorbox{Employer!30}{Model}\colorbox{Worker!30}{Data} $\bigstar$ \\ \hline

    \end{tabular}
    \vspace{-5mm}
\end{table*}

\subsection{How to Design ML Microtasks}
\label{sec:how2design}
When we come to the scenario of contract-based FL, we first need to define how to design ML microtasks for crowd workers. 
Unfortunately, previous collaborative ML studies~\cite{li2021survey, nguyen2021federated} rarely discuss this question because they default to assuming that all employers or workers have the same resource type to fit into their modeling frameworks.
On the contrary, in contract-based FL, we leave the freedom of design microtasks and choosing modeling method to the employers.

%The answer to this question depends on what resources the employers have and what tasks the workers can perform.
The answer depends on what resources the employers have and what tasks the workers can perform.
For example, in horizontal FL, the server should provide an initial model, and clients should have training data under the same feature space and provide computational power. However, in vertical FL, the clients should have training data under the same sample space (i.e., ID space). 
This difference determines which modeling framework employers can adopt. 
%Following this idea, we provide comparisons of typical FL~\cite{mcmahan2017communication, wang2023data, vepakomma2019split, kuang2023federatedscope, qin2023federated, berdoz2022scalable}, distributed ML~\cite{ryabinin2021moshpit, atre2021distributed, yuan2022decentralized, aketi2024cross, tan2022fedproto, michieli2021prototype, borzunov2022training}, and blockchain-based~\cite{ryabinin2020towards, mcconaghy2022ocean} modeling frameworks in TABLE~\ref{tab:train}.
Following this idea, we provide comparisons of typical FL~\cite{mcmahan2017communication}, distributed ML~\cite{ryabinin2021moshpit}, and blockchain-based~\cite{mcconaghy2022ocean} modeling frameworks in TABLE~\ref{tab:train}.
For simplicity, we introduce the \textbf{Platform} between Employer and Worker in some frameworks to make it adaptable for contract-based FL. 
You can quickly restore the original structure by merging Employer and Platform columns.
To represent the transmission of resources, we mark resources provided by each group of entities with different colors.
The computational resource used for training is marked with $\bigstar$.
By this way, we can conveniently find the feasible and privacy-preserving modeling frameworks for collaborative learning. 
Here, we briefly introduce these frameworks group by scenarios\footnote{From the viewpoint of Employers.}.

\subsubsection{Train from scratch with workers' private data}
There are four frameworks that support this scenario. 
The first is \textbf{Centralized ML}, which requires workers to upload their data to a cloud platform for modeling. However, this approach is not feasible for FL scenarios due to privacy concerns. 
The second is \textbf{FedAvg}~\cite{mcmahan2017communication}, which requires workers to upload computed gradients upon the scratch model and their private data (horizontally split). Computing resources need to be provided by workers as well.
The third is \textbf{SplitNN}~\cite{vepakomma2019split}, where each worker trains the cut layers of the model using their vertically split data. 
The hidden representations from these layers are uploaded to the platform for the training of the remaining layers.
The last and distinctive is a Web3 system named \textbf{Ocean}~\cite{mcconaghy2022ocean}.
It enables workers to upload only the metadata of their local dataset to a blockchain-based platform to attract buyers. 
Interested buyers then need to purchase datatokens to access the local dataset and deploy their modeling algorithm. 
To ensure the privacy and IP security of workers, only trusted algorithms are allowed, and only model predictions will be sent to buyers.

\subsubsection{Train from FMs}
FMs, trained on larger-scale data with robust generalization abilities across various downstream tasks, serve as a solid foundation for collaborative training. 
Yuan \textit{et al.} proposed \textbf{DT-FM}~\cite{yuan2022decentralized} to establish an effective geo-distributed learning system across 8 regions for training the language model GPT-3.
Similarly, Borzunov \textit{et al.}\cite{borzunov2022training} recruited volunteer nodes to train a transformer model over the Internet. 
To enhance system robustness, \textbf{Moshpit SGD}~\cite{ryabinin2021moshpit} divides worker nodes into small, independent groups to ensure that all-reduce is not affected by the failure of a single participant.
Additionally, \textbf{VC-ASGD}\cite{atre2021distributed} divides the deep learning training job into asynchronous parameter update subtasks to improve scalability.
In these cases, employers are required to supply the initial model and training data needed to launch training subtasks. 
This training paradigm is particularly suitable for worker nodes that possess computational and communication resources but lack their own training data.

Instead of initiating training from scratch and consuming substantial computational resources, we can also collaboratively adapt these FMs to specific local tasks based on relatively restricted hardware and data size through fine-tuning.
Recently, Woisetschl{\"a}ger \textit{et al.}~\cite{woisetschlager2023federated} have explored the opportunity to fine-tune large language models, such as FLAN-T5, by edge devices.
Simultaneously, \textbf{FS-LLM}~\cite{kuang2023federatedscope} employs Parameter-Efficient Fine-Tuning (PEFT) methods for federated fine-tuning of LLaMA-7B, minimizing communication and computation costs.
\textbf{FATE-LLM}~\cite{fan2023fate} offers a solution called FedHeteroLLM, leveraging KD to train a mentee model from its local pre-trained LLM for federated aggregation.
To alleviate the computational burden associated with backpropagation-based optimization methods, \textbf{FedKSeed}~\cite{qin2023federated} employs zeroth-order optimization (ZOO). 
Only seed and scalar gradients need to be transmitted for federated fine-tuning.

\subsubsection{Ensemble workers' private model} %优点：支持模型异构
As we presented in \S\ref{sec:taxonomy}, ensembling is an effective method to integrate knowledge, and the distributed version is proposed by \textbf{DMoE}~\cite{ryabinin2020towards}. 
It employs a Distributed Hash Table (DHT) to store metadata and worker statuses, constructing a decentralized expert network. 
To make inferences using this network, each model user must train a local gating network to select a subset of experts tailored to their input.
No additional training required, Wang \textit{et al.} purposed \textbf{DeDES}~\cite{wang2023data}, which selects a diverse subsect of weak models from population and make inference by voting. 
The unique advantage of ensembling lies in its inherent support for model heterogeneity, and its integration is extensible. 
This capability enables the establishment of a cooperative network while maintaining flexibility and availability.

\subsubsection{Aggregated workers's private representations}
Another popular scheme for organizing ML microtasks involves sharing the learned representations of workers. 
For instance, the use of class-conditional average of last hidden layer activations (aka. Prototypes) can enhance class discrimination across different clients (e.g., \textbf{FedProto}~\cite{michieli2021prototype, tan2022fedproto} and \cite{berdoz2022scalable}).
With no central server required, \textbf{CCL}~\cite{aketi2024cross} presents a decentralized learning approach based on cross-workers prototypes sharing.
To configure the microtasks, employers only need to set the target class (no scratch model reqiured), and the platform can then collect and distribute per-class prototypes among workers.
%However, the choosing between central solution and decentral solution is depends on what kind of infrastructure are perfered.
%For example, we can decentralized the store of local mode by IPFS~\cite{benet2014ipfs} protocol

\subsection{Limitations of Contract-based FL}
\label{sec:limitations_cbfl}
In contract-based FL systems, employers can publish their personalized ML crowdsourcing tasks, considering both their local resources and the resources of the target clients. 
However, to meet the flexibility requirements of widely supporting training methods, the contract platform should offer highly compatible APIs that can integrate various distributed training frameworks. 
Additionally, it needs to provide comprehensive audit against malicious users and plagiarism, significantly increasing the complexity compared to traditional FL and query-based FL.
Meanwhile, to establish confidence among users, a fair and consensual third-party model evaluation mechanism should be established, which, however, is almost unexplored in the current academic studies.

\section{Conclusion}
\label{sec:conclusion}
Traditional federated learning systems with a server-dominated cooperation framework suppress the enthusiasm of participants and limit the further extension of such collaboration. 
To explore the opportunity to establish a more open and reciprocal cooperation platform, we investigate current progress in federated learning, decentralized machine learning, and model reusing systems. 
In this way, we depict two rough sketches of open federated learning platforms: query-based federated learning paltform and contract-based federated learning paltform. 
Based on these two proposed platforms, we survey their possible supported techniques and their related legal issues, including ML licensing and copyrightability. 
We believe this survey can encourage a rethinking of current collaborative ML systems design and lead to the pervasive availability of AI for everyone.

%Therefore, we don't delve into this topic in this survey.

% 目标：1）保护任务发布方的利益和隐私；2）保护数据所有者的利益和隐私；3）遏制侵权行为

% 如何衡量模型质量：1）第三方evaluation

% 如果不采用pipeline的方式训练，如何拆分训练任务？

% 如何发现可用的FL节点
% 几种训练组织方式

\section*{Acknowledgments}
This research is supported by the National Research Foundation Singapore and DSO National Laboratories under the AI Singapore Programme (AISG Award No: AISG2-RP-2020-018). 
Any opinions, findings and conclusions or recommendations expressed in this material are those of the authors and do not reflect the views of National Research Foundation, Singapore.

\begin{comment}
{\appendix[Proof of the Zonklar Equations]
Use $\backslash${\tt{appendix}} if you have a single appendix:
Do not use $\backslash${\tt{section}} anymore after $\backslash${\tt{appendix}}, only $\backslash${\tt{section*}}.
If you have multiple appendixes use $\backslash${\tt{appendices}} then use $\backslash${\tt{section}} to start each appendix.
You must declare a $\backslash${\tt{section}} before using any $\backslash${\tt{subsection}} or using $\backslash${\tt{label}} ($\backslash${\tt{appendices}} by itself
starts a section numbered zero.)}

%{\appendices
%\section*{Proof of the First Zonklar Equation}
%Appendix one text goes here.
% You can choose not to have a title for an appendix if you want by leaving the argument blank
%\section*{Proof of the Second Zonklar Equation}
%Appendix two text goes here.}
\end{comment}

\bibliographystyle{IEEEtran}
\bibliography{IEEEabrv,REF}

\begin{comment}
\section{Biography Section}
If you have an EPS/PDF photo (graphicx package needed), extra braces are
 needed around the contents of the optional argument to biography to prevent
 the LaTeX parser from getting confused when it sees the complicated
 $\backslash${\tt{includegraphics}} command within an optional argument. (You can create
 your own custom macro containing the $\backslash${\tt{includegraphics}} command to make things
 simpler here.)
 
\vspace{11pt}

\bf{If you include a photo:}\vspace{-33pt}
\end{comment}

\begin{IEEEbiography}[{\includegraphics[width=1in,height=1.25in,clip,keepaspectratio]{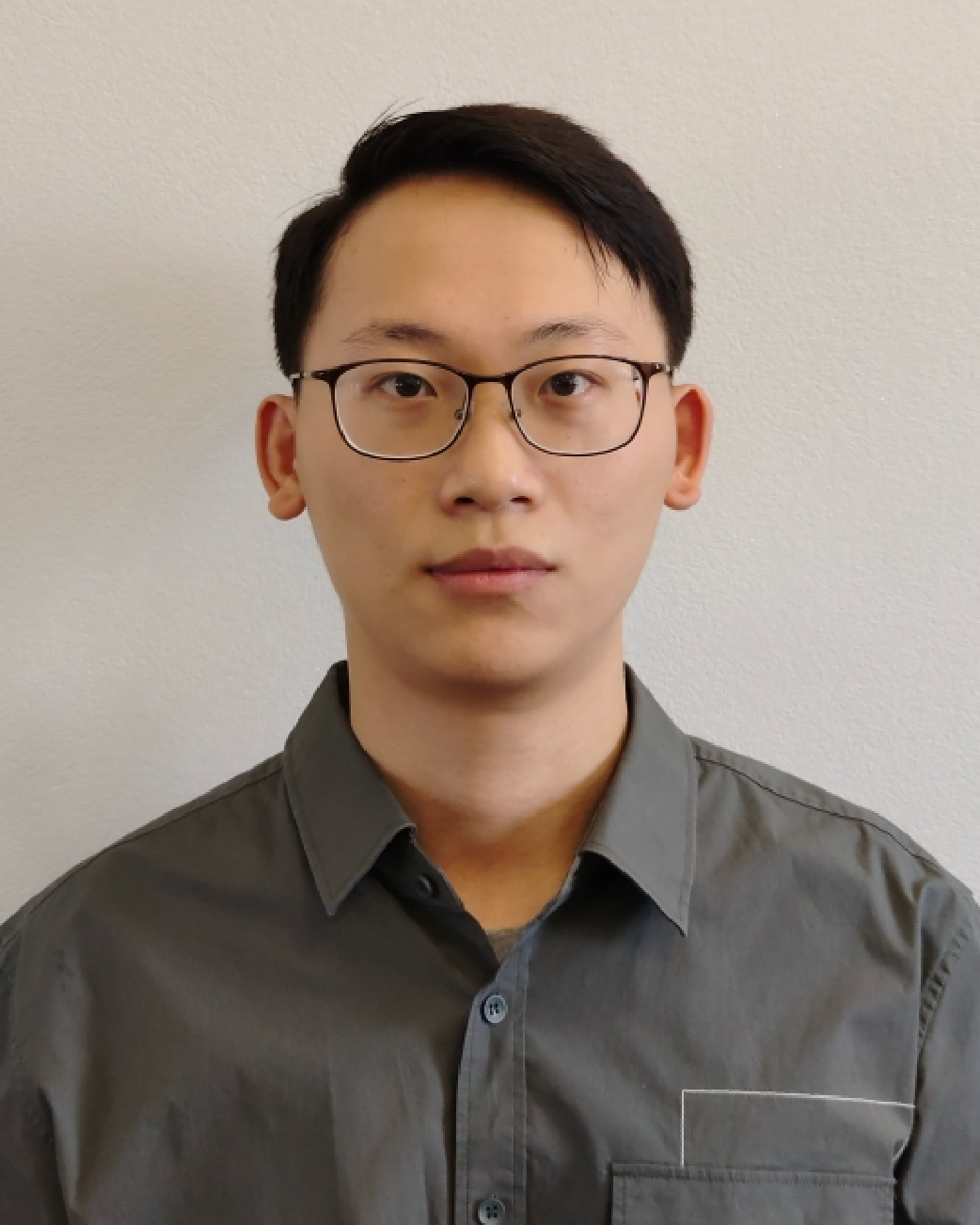}}]{Moming Duan}
is a Research Fellow at Institute of Data Science, National University of Singapore. He received PhD degree in computer science from Chongqing University (2017-2022). He has published several papers in prestigious conferences and journals including TPDS, WWW, ICCD, Neural Networks, TMLR. His research interests include Federated Learning, Collaborative Machine Learning, and AI Licensing.
\end{IEEEbiography}

%\vspace{11pt}

\begin{IEEEbiography}[{\includegraphics[width=1in,height=1.25in,clip,keepaspectratio]{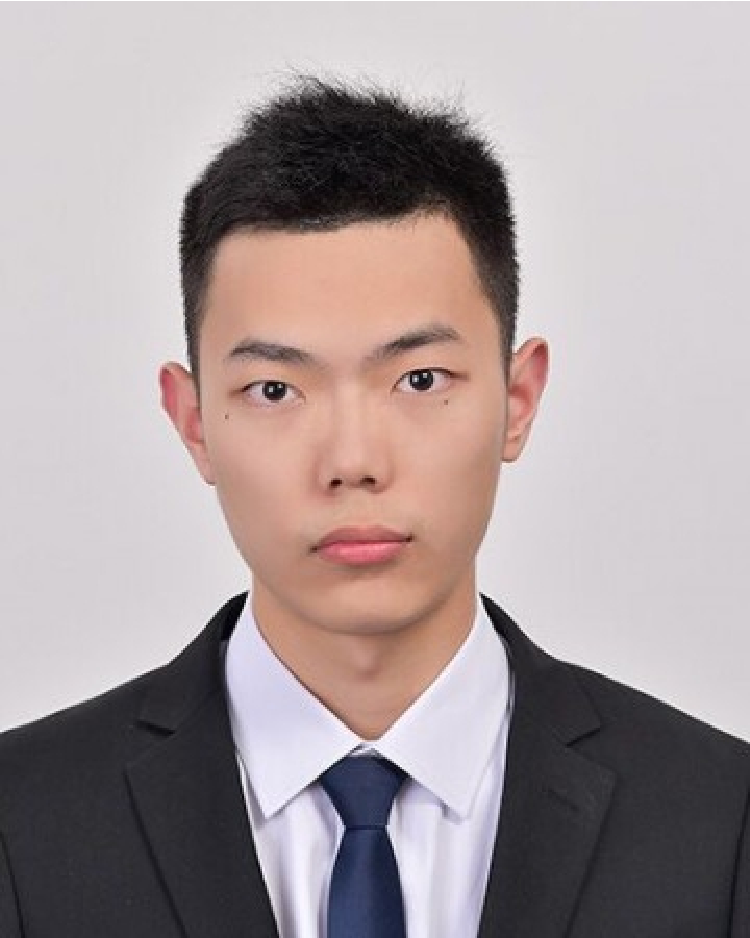}}]{Qinbin Li}
is a postdoc at UC Berkeley. He received his PhD degree from National University of Singapore. He is a recipient of Google PhD Fellowship 2021. His research interests include federated learning, trustworthy machine learning, and systems.
\end{IEEEbiography}

\begin{IEEEbiography}[{\includegraphics[width=1in,height=1.25in,clip,keepaspectratio]{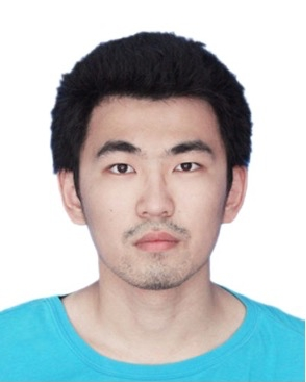}}]{Linshan Jiang}
is currently a Research Fellow at Institute of Data Science, National University of Singapore. He obtained his Ph.D. degree in computer science and engineering from Nanyang Technological University, Singapore in 2022. He has published several papers on the top conference and journals in CPS-IoT. His research interests focus on the privacy and security in the distributed AI system, including federated/collaborative learning, blockchain-enabled AI and resilient AIoT system.
\end{IEEEbiography}

\begin{IEEEbiography}[{\includegraphics[width=1in,height=1.25in,clip,keepaspectratio]{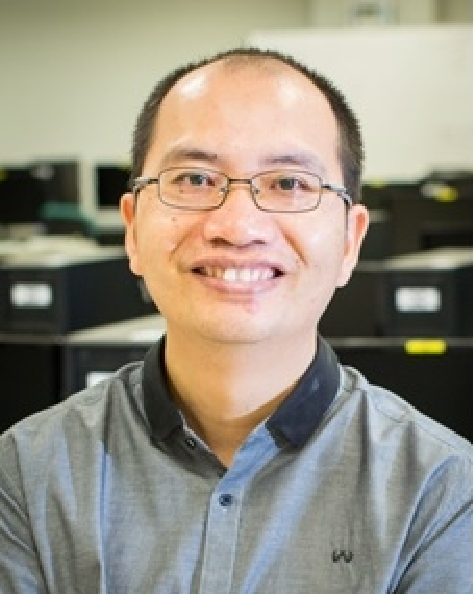}}]{Bingsheng He}
received the bachelor degree in computer science from Shanghai Jiao Tong University (1999-2003), and the PhD degree in computer science in Hong Kong University of Science and Technology (2003-2008). He is a Professor in School of Computing, National University of Singapore. His research interests are high performance computing, distributed and parallel systems, and database systems.
\end{IEEEbiography}

\vfill

\end{document}